# Strain-Induced Polarization Enhancement in BaTiO$_3$ Core-Shell Nanoparticles

Eugene A. Eliseev[1], Anna N. Morozovska[2], Sergei V. Kalinin[3*], and Dean R. Evans[4†]

[1] Institute for Problems of Materials Science, National Academy of Sciences of Ukraine,
Krjijanovskogo 3, 03142 Kyiv, Ukraine

[2] Institute of Physics, National Academy of Sciences of Ukraine,
46, pr. Nauky, 03028 Kyiv, Ukraine

[3] Department of Materials Science and Engineering, University of Tennessee,
Knoxville, TN, 37996, USA

[4] Air Force Research Laboratory, Materials and Manufacturing Directorate, Wright-Patterson Air Force Base, Ohio, 45433, USA

**Abstract**

Despite fascinating experimental results, the influence of defects and elastic strains on the physical state of nanosized ferroelectrics is still poorly explored theoretically. One of unresolved theoretical problems is the analytical description of the strongly enhanced spontaneous polarization, piezoelectric response, and dielectric properties of ferroelectric oxide thin films and core-shell nanoparticles induced by elastic strains and stresses. In particular, the 10-nm quasi-spherical BaTiO$_3$ core-shell nanoparticles reveal a giant spontaneous polarization up to 130 μC/cm$^2$, where the physical origin is a large Ti off-centering. The available theoretical description cannot explain the giant spontaneous polarization observed in these spherical nanoparticles. This work analyzes polar properties of BaTiO$_3$ core-shell spherical nanoparticles using the Landau-Ginzburg-Devonshire approach, which considers the nonlinear electrostriction coupling and large Vegard strains in the shell. We reveal that a spontaneous polarization greater than 50 μC/cm$^2$ can be stable in a (10-100) nm BaTiO$_3$ core at room temperature, where a 5 nm paraelectric shell is stretched by (3-6)% due to Vegard strains, which contribute to the elastic mismatch at the core-shell interface. The polarization value 50 μC/cm$^2$ corresponds to high tetragonality ratios (1.02 - 1.04), which is further increased up to 100 μC/cm$^2$ by higher Vegard strains and/or intrinsic surface stresses leading to unphysically high tetragonality ratios (1.08 - 1.16). The nonlinear electrostriction coupling and the elastic mismatch at the core-shell interface are key physical factors of the spontaneous polarization enhancement in the core. Doping with the highly-polarized core-shell nanoparticles can be useful in optoelectronics and nonlinear optics to increase beam coupling efficiency, electric field enhancement, reduced switching voltages, ionic contamination elimination, catalysis, and electrocaloric nanocoolers.

---

[*] Corresponding author: sergei2@utk.edu

[†] Corresponding author: dean.evans@afrl.af.mil

## I. INTRODUCTION

Despite fascinating experimental results, the influence of size and screening effects, defects, and elastic strains on the physical state of nanosized ferroelectrics is poorly explored theoretically. This precludes the fundamental understanding of underlying physical mechanisms and significantly limits practical applications of these nanomaterials [1, 2]. One of the unresolved theoretical problems is the analytical description of the strongly enhanced spontaneous polarization, piezoelectric response, and dielectric properties of ferroelectric oxide thin films [3, 4] and core-shell nanoparticles [5, 6, 7] induced by elastic strains (e.g., by strains created by a pure lattice mismatch) and/or stresses created by injected elastic defects (e.g., Vegard strains). A solution to this problem would allow one to achieve significant progress in advanced applications of these nanosized ferroelectrics.

Highly enhanced ferroelectricity can be reached in (Hf,Zr)O$_2$ thin films by light-ion bombardment [3]. Due to the ion bombardment, the massive formation of oxygen vacancies appears in the film. These vacancies form elastic dipoles which, due to the mechanism of electrostriction, enhance the spontaneous electric polarization and local piezoelectric response of the film. In a similar way, oxygen-vacancy injection is a pathway to a strong enhancement of the electromechanical response in BaTiO$_3$ thin films [4]. The electromigration and diffusion of the oxygen vacancies, which are elastic dipoles, can induce relatively strong mechanical stresses in ABO$_3$-type perovskites (so-called Vegard effect) [8, 9] and Fe centers recharging in photorefractive oxide ferroelectrics [10]. Due to the electrostriction coupling, the Vegard effect increases the temperature of the paraelectric-ferroelectric phase transition up to 440 K, which is significantly higher than the bulk transition temperature, $T_C = 391$ K. The tetragonality ratio, $c/a$, increases up to 1.032 near the film surface and decreases to the bulk value ($c/a = 1.011$) only at a distance of 20 lattice constants from the surface. The local piezoelectric response increases a factor of four at room temperature. Since the local piezoelectric response is regarded to be proportional to the ferroelectric polarization, it can reach 100 μC/cm$^2$ at 300 K in the BaTiO$_3$ thin films [4].

A decade earlier, Zhu et al. [5] observed a sharp increase of the $c/a$ ratio up to 1.055 in (5-10) nm BaTiO$_3$ nanospheres (i.e., no shell) synthesized using solvothermal methods (in comparison with $c/a \approx 1.002$ in 20-nm nanospheres) at 293 K. They concluded the appearance of a reentrant tetragonal phase for particle sizes less than 20 nm was due to competition of shear and compressive stresses on the particle surface.

The 10-nm quasi-spherical BaTiO$_3$ core-shell nanoparticles reveal a giant spontaneous polarization up to 130 µC/cm$^2$ at room temperature, which is five times greater than the bulk value 26 µC/cm$^2$ (see Refs. [6, 11, 12] and references therein). The incorporation of these nanoparticles can have multiple benefits in various applications, such as enhanced beam coupling efficiency [13], reduced switching voltages/DC bias [14, 15], ionic contamination elimination [16], and catalysis [17]. Experimental realizations of quasi-spherical BaTiO$_3$ ferroelectric nanoparticles are abundant, and the sizes of (5 - 50) nm are typical experimental values [18, 19, 20, 21]. The nanoparticles embedded in heptane and oleic acid produce core-shell nanoparticles, where the oleic acid is transformed into an organic crystalline (metal carboxylate) shell surrounding the inorganic BaTiO$_3$ core resulting from mechanochemical synthesis during the ball-milling process [22]. The metal carboxylate coating/shell around the BaTiO$_3$ core can be in two forms – one is crystalline and provides a lattice mismatch at the core-shell interface, and the other is non-crystalline without mismatch conditions [6]. The observed polarization enhancement is possible for the BaTiO$_3$ core with the crystalline shell.

It is worth noting that ferroelectric nanoparticles with such giant spontaneous polarization values are not always achievable from the ball-milled mixture without additional processing. Typically, the harvesting technique described in Ref. [23] is required, which relies on an electric field gradient to selectively harvest ferroelectric nanoparticles with the strongest dipole moments from bulk nanoparticle ball-milled mixtures. The total nanoparticle yield of strong dipoles using the ball-milling/mechanochemical synthesis technique has varied from nearly 0% to ∼100% [23], while the harvesting technique has shown to repeatedly provide 100% usable strong dipoles from these mixtures. The harvesting technique was described by Yu. Reznikov in Ref. [24] as being a "break-through" in solving the problem of irreproducibility. Alternatively, a process of separation via centrifuge or simple sedimentation, where larger particles drop and form agglomerates and small particles remain in suspension, has also shown to provide strong dipoles. Although this latter method has provided the strong dipole particles used in the Ref. [6, 7], its effectiveness and reliability compared to the proven harvesting technique has not been determined.

The physical origin of the giant spontaneous polarization in the quasi-spherical BaTiO$_3$ core-shell nanoparticles remained a mystery for a long time, until recent X-ray spectroscopic measurements [7] revealed a large Ti-cation off-centering in 10-nm nanoparticles near 300 K confirmed by the tetragonality ratio $c/a \approx 1.0108$ (in comparison with $c/a \approx 1.0075$ for 50 nm nanoparticles). The off-centering of Ti-cations is a key factor in producing the enhanced spontaneous polarization in the nanoparticles. Sharp crystalline-type peaks in the barium oleate Raman spectra suggest that this component in the composite core-shell matrix, a product of mechanochemical synthesis, stabilizes an enhanced polar structural phase of the BaTiO$_3$ core.

To the best of our knowledge, there is no available theoretical description that can explain the giant spontaneous polarization repeatedly observed in the BaTiO$_3$ core-shell nanoparticles. Indeed, the surface bond contraction mechanism can only decrease the polarization of the ferroelectric ABO$_3$-type perovskite nanoparticles [25, 26]. Some theoretical papers [27, 28, 29] predict the enhancement of a reversible spontaneous polarization in prolate nanoellipsoids, nanorods, and nanowires of ABO$_3$-type perovskites, when their polarization is directed along the longest axis. A significant polarization enhancement can also appear due to the high positive surface tension coefficient and negative linear electrostriction coupling coefficients $Q_{12}$; the dependence of the Curie temperature on the particle radius $R$ is proportional to the positive value $-\frac{4\mu}{R}Q_{12}$ (see e.g., Table 1 in Ref. [30]). Furthermore, this same mechanism leads to a significant reduction of the Curie temperature $T_C$ and spontaneous polarization in spherical BaTiO$_3$ nanoparticles specifically, because the value $-\frac{2\mu}{R}(2Q_{12} + Q_{11})$ is negative, since the condition $\mu > 0$ is required for the surface equilibrium and $2Q_{12} + Q_{11} > 0$ for BaTiO$_3$. The flexo-chemical effect [31] emerging from the joint action of the Vegard stresses and flexoelectric effect, can increase $T_C$, spontaneous polarization, and $c/a$ in ultra-small (5 nm or less) spherical BaTiO$_3$ nanoparticles and explain experimental results [5], although the effect rapidly disappears with a radius increase ($\sim \frac{1}{R^2}$) and requires very high values of the flexoelectric coupling and intrinsic strains. The Vegard strains ($w_{ij}^s$) can significantly increase $T_C$ in spherical KTa$_{1-x}$Nb$_x$O$_3$ nanoparticles with $R < 30$ nm [30], as well as in the BaTiO$_3$ core-shell nanoparticles with $R < 10$ nm (see Fig. 9 in Ref. [32]); however, the influence of strong Vegard strains (i.e., $w_{ij}^s$, >0.5%) on the BaTiO$_3$ spontaneous polarization was not studied in Ref. [32]. Nonlinear electrostriction coupling, which needs to be considered for strains higher than 1%, induces an instability of the 6-th order BaTiO$_3$ thermodynamic potential, because a higher strain changes the positive sign of the 6-th order polarization term at temperatures well below 350 K. Note, the nonlinear electrostriction coupling can be very important for a correct description of polar properties of strained ferroelectric thin films [33, 34] and core-shell nanoparticles [35, 36].

All aforementioned and many other theoretical works considering BaTiO$_3$ nanoparticles are based on the Landau-Ginzburg-Devonshire (**LGD**) phenomenological approach, which includes the 2-nd, 4-th, and 6-th powers of polarization in the LGD free energy expansion and only considers linear electrostriction coupling (see e.g., Ref. [32] and references therein). This work analyzes polar properties of core-shell BaTiO$_3$ nanoparticles using LGD free energy functional proposed by Wang et al. [37], which includes the 8-th power of polarization, and thus allows high Vegard strains in the shell and the nonlinear electrostriction coupling in the core to be considered.

## II. PROBLEM FORMULATION, MAIN IDEAS, AND CALCULATION DETAILS

### A. The problem formulation

Let us consider a spherical BaTiO$_3$ core-shell nanoparticle in the tetragonal phase, whose core of radius $R_c$ is a single-domain with a spontaneous polarization $\vec{P}_s$ directed along one of the crystallographic directions (e.g., along the polar axis X$_3$). The crystalline core has a perfect structure (without any defects) and is considered to be insulating (without any free charges). The core is covered with a crystalline shell of thickness $\Delta R$ and outer radius $R_s$. The shell is semiconducting and paraelectric due to the high concentration of free charges and elastic defects. The free charges provide an effective screening of the core spontaneous polarization and prevent domain formation. The elastic defects induce strong Vegard strains, $w_{ij}^s$, which are regarded as cubic, $w_{ij}^s = \delta_{ij} w_s$, where $\delta_{ij}$ is the Kronecker-delta symbol and $w_s$ is the magnitude of Vegard strains in the shell, where as a rule, cannot exceed (5 – 10)%. These strains can stress the core due to the elastic mismatch at the core-shell interface. The effective screening length in the shell, $\lambda$, is small (less than 1 nm), and its relative dielectric permittivity tensor, $\varepsilon_{ij}^s$, is isotropic, $\varepsilon_{ij}^s = \delta_{ij} \varepsilon_s$, which can be very high ($10^2 – 10^3$) as anticipated for the paraelectric state. The core-shell nanoparticle is placed in a dielectric medium (polymer, gas, liquid, air, or vacuum) with an effective dielectric permittivity, $\varepsilon_e$. The core-shell geometry is shown in **Fig. 1**.

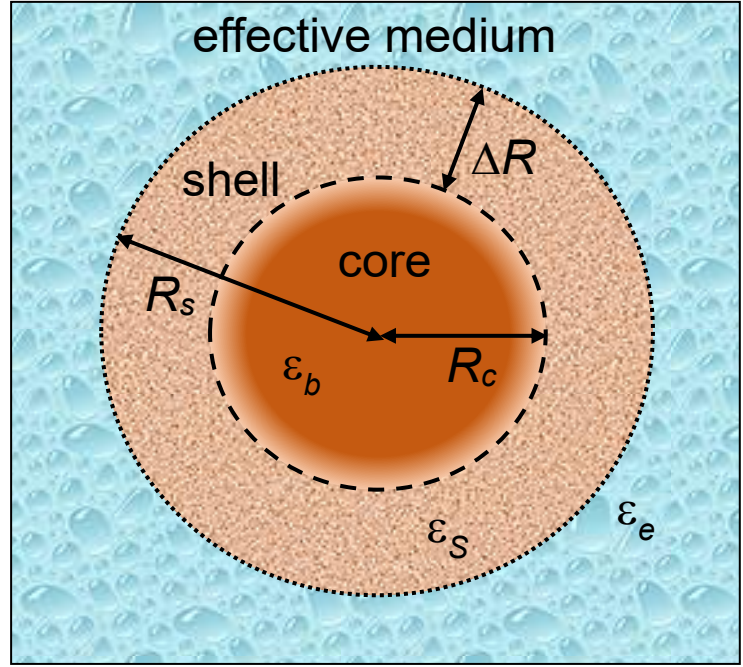

**FIGURE 1.** A spherical core-shell nanoparticle: a ferroelectric core of radius $R_c$ is covered with a paraelectric shell of thickness $\Delta R$, which is full of elastic defects and free charges. The nanoparticle is placed in an isotropic dielectric medium; $\varepsilon_b$, $\varepsilon_s$, and $\varepsilon_e$ are the core background, shell, and surrounding media dielectric permittivities.

The LGD free energy density includes the Landau-Devonshire expansion in even powers of the polarization $P_3$ (up to the 8-th power), the Ginzburg gradient energy, and the elastic and electrostriction energies, which are listed in **Appendix A1** of **Supplementary Materials** [38]. The equilibrium polarization distribution in the core follows from the Euler-Lagrange equation, which in turn follows from the minimization of the LGD free energy, and has the form:

$$[\alpha - 2\sigma_i(Q_{i3} + W_{ij3}\sigma_j)]P_3 + (\beta - 4Z_{i33}\sigma_i)P_3^3 + \gamma P_3^5 + \delta P_3^7 - g_{33kl}\frac{\partial^2 P_3}{\partial x_k \partial x_l} = E_3. \quad (1)$$

Here the parameters $\alpha$, $\beta$, $\gamma$, and $\delta$ are the 2-nd, 4-th, 6-th, and 8-th order Landau expansion coefficients in the $P_3$-powers of the free energy corresponding to the bulk BaTiO$_3$. The values $\sigma_i$ denote diagonal components of a stress tensor in Voigt notation, and subscripts $i, j = 1 - 6$. The values $Q_{i3}$, $Z_{i33}$, and $W_{ij3}$ are the components of a single linear and two nonlinear electrostriction strain tensors in Voigt notation, respectively [39, 40]. The values $g_{33kl}$ are polarization gradient coefficients in matrix notation, and subscripts $k, l = 1 - 3$. The Neumann boundary condition for $P_3$ at the nanoparticle surface S is $g_{33kl} n_k \frac{\partial P_3}{\partial x_l}\Big|_S = 0$, where $\vec{n}$ is the outer normal to the surface. These conditions are also called "natural", because corresponding surface energy is zero in the case.

The value $E_3$ is an electric field component, co-linear with the polarization $P_3$, which is a superposition of external and depolarization fields, $E_3^0$ and $E_3^e$, respectively. The quasi-static field $E_3$ is related to the electric potential $\varphi$ as $E_3 = -\frac{\partial \varphi}{\partial x_3}$. The potential $\varphi$ satisfies the Poisson equation inside the particle core, the Debye equation in the semiconducting shell, and the Laplace equation outside the screening shell (see **Appendix A1** for details).

The parameters of the BaTiO$_3$ core and paraelectric shell used in our calculations are listed in **Tables AI** and **AII** in **Appendix A1**, respectively. In order to focus on the influence of linear and nonlinear electrostriction effects, we do not consider the surface tension and flexoelectric coupling in this work. The scalar parameters $\alpha$, $\beta$, $\gamma$, and $\delta$ in Eq.(1) are related with tensorial coefficients, $a_i$, $a_{ij}$, $a_{ijk}$, and $a_{ijkl}$ in the LGD free energy (A.2) in a conventional way. Namely, $\alpha = 2a_1$, $\beta = 4a_{11}$, $\gamma = 6a_{111}$, and $\delta = 8a_{1111}$ in the tetragonal phase of the core. In what follows, we consider two sets of the BaTiO$_3$ core parameters taken from Wang et al. [37]:

1) A widely used "**2-4-6 LGD**" free energy functional, which includes the 2-nd, 4-th, and 6-th powers of the polarization $P_3$ in the Landau-Devonshire free energy with the nonlinear electrostriction coupling tensors set to zero, $Z_{ijk} = 0$, $W_{ijk} = 0$ (see the second column in **Table AI**). The coefficient $\alpha$ depends linearly on the temperature $T$, $\alpha(T) = \alpha_T(T - T_C)$, where $T_C = 381$ K is the Curie temperature. Also, the coefficients $\beta$ and $\gamma$ linearly depend on the temperature and can change their sign. Since $\delta = 0$, the 2-4-6 LGD functional becomes unstable above 445 K, when $\gamma$ becomes negative. This is very inconvenient for the modeling of strongly stressed nanoparticles, because elastic stresses above 1% can reduce the instability temperature (room or lower), making the 2-4-6 LGD free energy functional unsuitable for the modeling of strongly stressed nanoparticles.

2) More rarely used is the "**2-4-6-8 LGD**" free energy functional, which includes the 2-nd, 4-th, 6-th, and 8-th powers of the polarization $P_3$ in the Landau-Devonshire free energy without consideration of the nonlinear electrostriction coupling effect, i.e., $Z_{ijk} = 0$ and $W_{ijk} = 0$ (see the third column in **Table AI**). For this case, the coefficient $\alpha$ also depends linearly on the temperature $T$, $\alpha(T) = \alpha_T(T - T_C)$, where $T_C = 391$ K. The coefficients $\beta$ and $\gamma$ linearly depend on the temperature and can change sign, but the coefficient $\delta$ is positive and temperature-independent. Since $\delta > 0$, the 2-4-6-8 LGD free energy functional is stable for arbitrary temperatures, and thus is suitable for artifact-free modeling of strongly stressed nanoparticles.

The most important part of this work is to study how the nonlinear electrostriction coupling effect ($Z_{ijk} \neq 0$ and $W_{ijk} \neq 0$) influences the stability conditions of the 2-4-6 and 2-4-6-8 LGD free energy functionals, and determine the changes in spontaneous polarization that are induced by the nonlinear electrostriction effect. Different sets of electrostriction coefficients $Z_{ijk}$ determined from different experiments or ab initio calculations are listed in **Table AIII**. In accordance with the table, the values of $Z_{ijk}$ can vary from -14.2 m$^8$/C$^4$ to +0.38 m$^8$/C$^4$ in a bulk BaTiO$_3$; and the "scattering" range is a very wide. Since recent works reveal an extraordinarily high electrostriction due to the interface effects [41], we can assume that the possible range of $Z_{ijk}$ variation can be even wider in the core-shell BaTiO$_3$ nanoparticles. These speculations give us some grounds to vary $Z_{ijk}$ within the range from -1.5 m$^8$/C$^4$ to +1.5 m$^8$/C$^4$ to look for optimal values that correspond to the highest spontaneous polarization and the best related properties.

### B. Approximate analytical description

For the case of natural boundary conditions used in this work, $g_{33ij}n_i \frac{\partial P_3}{\partial x_j} = 0$, small $\lambda$, and relatively large gradient coefficients $|g_{ijkl}| > 10^{-11}$ C$^{-2}$m$^3$J, the polarization gradient effects can be neglected in a single-domain state, which reveals to have a minimal energy in comparison to polydomain states. The field dependence of a quasi-static single-domain polarization can be found from the following equation:

$$\alpha^* P_3 + \beta^* P_3^3 + \gamma P_3^5 + \delta P_3^7 = E_3^e. \qquad (2)$$

The depolarization field, $E_3^d$, and stresses, $\sigma_i$, contribute to the "renormalization" of coefficient $\alpha(T)$, which becomes the temperature-, radius-, stress-, and screening length- dependent function $\alpha^*$ [36]:

$$\alpha^*(T, R_c, \sigma_i) = \alpha(T) + \frac{1}{\varepsilon_0(\varepsilon_b + 2\varepsilon_s + R_c/\lambda)} - \sigma_i(2Q_{i3} + W_{ij3}\sigma_j). \tag{3a}$$

The derivation of the second term in Eq.(3a) is given in Ref. [42]. Here, $\lambda = \lambda(E_3^d)$ can be a rather small value (less than 0.1 – 1 nm) due to free charges and surface band bending in the shell. Resulting from the nonlinear electrostriction coupling, the coefficient $\beta^*$ is "renormalized" by elastic stresses as

$$\beta^*(\sigma_i) = \beta - 4Z_{i33}\sigma_i. \tag{3b}$$

In the right-hand side of Eq.(2), $E_3^e$ is the static external field inside the core, for which the estimate $E_3^e \approx \frac{3\varepsilon_s E_3^0}{\varepsilon_b + 2\varepsilon_s + R_c/\lambda}$ is valid. If $\lambda(E_3^0) \gg R_c$ and $\varepsilon_s \sim \varepsilon_b$, the field in the core is of the same order as the applied field $E_3^0$ (see details in Ref. [36]).

Elastic stresses in the core, $\sigma_i$, induced by the Vegard strains in the shell, can be calculated analytically using the method of successive approximations. When the spontaneous polarization is absent in the paraelectric phase of the core, or small in the "shallow" ferroelectric state located near the paraelectric-ferroelectric transition, the core can be considered as elastically-isotropic due to its cubic symmetry or very small tetragonality related with the small electrostriction contribution. In **Appendix A2** an approximate expression for the diagonal stresses in the core and shell is derived. The core stresses, further denoted as $\sigma_i^c$, are given by the expression:

$$\sigma_1^c = \sigma_2^c = \sigma_3^c = \frac{-2(R_s^3 - R_c^3)(Q_c P_3^2 + Z_c P_3^4 - w_s)}{2(R_s^3 - R_c^3)(s_{11}^c + 2s_{12}^c + W_c P_3^2) + R_s^3(s_{11}^s - s_{12}^s) + 2R_c^3(s_{11}^s + 2s_{12}^s)}. \tag{4a}$$

Here, $s_{ij}^c$ and $s_{ij}^s$ are the elastic compliances of the core and shell, respectively; $Q_c = (Q_{11} + 2Q_{12})/3$; $Z_c = (Z_{111} + 2Z_{211})/3$; and $W_c = (W_{111} + 2W_{112} + 2W_{123} + 4W_{122})/3$ are isotropic parts of the linear and nonlinear electrostriction tensors of the core, and $w_s$ is the Vegard strain in the shell. The electrostriction coupling can exist in the shell; however, it does not contribute to the solution (4a) for small $\lambda$, since the electric field is very small in the shell due to the high screening degree. The nondiagonal stresses are absent, $\sigma_4^c = \sigma_5^c = \sigma_6^c = 0$.

The corresponding free energy of the core-shell nanoparticle is:

$$G = \left[\alpha + \frac{1}{\varepsilon_0(\varepsilon_b + 2\varepsilon_s + R_c/\lambda)}\right]\frac{P_3^2}{2} + \frac{\beta P_3^4}{4} + \frac{\gamma P_3^6}{6} + \frac{\delta P_3^8}{8} + \frac{3(R_s^3 - R_c^3)(Q_c P_3^2 + Z_c P_3^4 - w_s)^2}{2(R_s^3 - R_c^3)(s_{11}^c + 2s_{12}^c + W_c P_3^2) + R_s^3(s_{11}^s - s_{12}^s) + 2R_c^3(s_{11}^s + 2s_{12}^s)}. \tag{4b}$$

Since inequality $s_{11}^s - 2s_{12}^s > 0$ follows from the mandatory condition of a positive quadratic-form of the elastic energy and $R_s^3 - R_c^3 > 0$, the denominators in Eq.(4) are always positive. The expressions (4) are accurate enough to provide a first approximation for the description of the core in a "deep" ferroelectric phase, when the absolute value of the polarization-dependent anisotropic contribution to the total strain is much smaller than other contributions (see **Appendix A2** for details). The approximation imposes definite conditions on poorly known (or unknown) anisotropic parts of the tensors $Z_{ijk}$ and $W_{ijk}$. In order to avoid complications, the tensors are regarded as isotropic. The condition $W_c \geq 0$ should be valid for the free energy stability at high $P_3$, and this condition is assumed hereinafter.

If the term $W_c P_3^2$ is small and positive, Eq.(4a) becomes much simpler for two important cases: 1) when elastic compliances of the core and the shell are the same: $s_{ij}^s = s_{ij}^c \equiv s_{ij}$, and 2) for shells with $\Delta R \ll R_c$. In these cases:

$$\sigma_1^c = \sigma_2^c = \sigma_3^c \approx \begin{cases} -\frac{2(R_s^3-R_c^3)}{3R_s^3} \frac{Q_c P_3^2 + Z_c P_3^4 - w_s}{s_{11}+s_{12}}, & s_{ij}^s = s_{ij}^c \equiv s_{ij}, \\ \frac{\Delta R}{R_s} \frac{1}{s_{11}^s + s_{12}^s} (w_s - Q_c P_3^2 - Z_c P_3^4), & \Delta R \ll R_c. \end{cases} \quad (5)$$

After substitution of the solution (5) in Eq.(2) and elementary transformations (see **Appendix A3** for details), the equation for polarization $P_3$ with renormalized coefficients $\alpha_R$, $\beta_R$, $\gamma_R$, $\delta_R$, and $\epsilon_R$ is derived as:

$$\alpha_R P_3 + \beta_R P_3^3 + \gamma_R P_3^5 + \delta_R P_3^7 + \epsilon_R P_3^9 = E_3^e. \quad (6a)$$

The renormalized coefficients are given by expressions:

$$\alpha_R = \alpha + \frac{1}{\varepsilon_0(\varepsilon_b + 2\varepsilon_s + R_c/\lambda)} - 4Q_c \frac{(R_s^3-R_c^3)w_s}{R_s^3(s_{11}+s_{12})} + \frac{4}{3}W_c \left(\frac{(R_s^3-R_c^3)w_s}{R_s^3(s_{11}+s_{12})}\right)^2, \quad (6b)$$

$$\beta_R = \beta + 6Q_c^2 \frac{2(R_s^3-R_c^3)}{3R_s^3(s_{11}+s_{12})} - 8Z_c \frac{(R_s^3-R_c^3)w_s}{R_s^3(s_{11}+s_{12})} - \frac{8}{3}W_c Q_c w_s \left(\frac{R_s^3-R_c^3}{R_s^3(s_{11}+s_{12})}\right)^2, \quad (6c)$$

$$\gamma_R = \gamma + \frac{12(R_s^3-R_c^3)}{R_s^3(s_{11}+s_{12})} Q_c Z_c + \frac{4}{3}W_c(Q_c^2 - 2Z_c w_s) \left(\frac{R_s^3-R_c^3}{R_s^3(s_{11}+s_{12})}\right)^2, \quad (6d)$$

$$\delta_R = \delta + Z_c^2 \frac{8(R_s^3-R_c^3)}{R_s^3(s_{11}+s_{12})} + \frac{8}{3}W_c Q_c Z_c \left(\frac{R_s^3-R_c^3}{R_s^3(s_{11}+s_{12})}\right)^2, \quad (6e)$$

$$\epsilon_R = \frac{4}{3}W_c Z_c^2 \left(\frac{R_s^3-R_c^3}{R_s^3(s_{11}+s_{12})}\right)^2. \quad (6f)$$

The renormalization of the coefficients in Eq.(6) is proportional to the ratio $\frac{R_s^3-R_c^3}{3R_s^3}$, which is close to $\frac{\Delta R}{R_s}$ for thin shells with $\Delta R \ll R_c$. The renormalization is most significant for small nanoparticles with $\Delta R \sim R_c$; it vanishes for $R_c \to R_s$ and is absent for thin films, where the curvature disappears, $R_c$ and $R_s$ tend to infinity, and their finite difference $\Delta R$ becomes the film thickness.

By definition, the tetragonality $c/a$ is proportional to the ratio:

$$\frac{c}{a} = \frac{1+u_3^c}{1+u_1^c}, \quad (7a)$$

where $u_3^c$ and $u_1^c$ are the core (denoted by the superscript $c$) strains written in Voigt notations. Because the core strains are small, $|u_3^c| \ll 1$ and $|u_1^c| \ll 1$, the tetragonality ratio is proportional to their difference, $\frac{c}{a} \approx 1 + u_3^c - u_1^c$. Since the tensors of nonlinear electrostriction coupling are assumed to be isotropic, their contribution to the $u_3^c$ and $u_1^c$ are the same, namely $u_3^c = C_1 + Q_{11}P_3^2$ and $u_1^c = C_1 + Q_{12}P_3^2$, where the function $C_1$ is given by Eq.(A.14a) in **Appendix A2**. Therefore, the deviation of the tetragonality ratio from unity is proportional to the anisotropy of linear electrostriction coefficients of the core:

$$\frac{c}{a} \approx 1 + (Q_{11} - Q_{12})P_3^2. \qquad (7b)$$

### III. RESULTS AND DISCUSSION

Numerical results presented in this section are obtained and visualized in Mathematica 13.2 [43]. The calculations are performed for a small screening length ($\lambda = 0.5$ nm), small thickness ($\Delta R = 5$ nm), and high dielectric permittivity ($\varepsilon_s = 500$) of the paraelectric shell. The spontaneous polarization, $P_s$, of a single-domain BaTiO$_3$ core is calculated for a range of the core radii $R_c = (5 - 50)$ nm as a function of both temperature, $T$, and Vegard strain, $w_s$, using 2-4-6 and 2-4-6-8 LGD free energies for zero (**Fig. 2**) and nonzero (**Fig. 3**) nonlinear electrostriction coupling tensors, $Z_{ijk}$ and $W_{ijk}$, which have different dimensionalities, are responsible for the coupling strength of different contributions in the free energy part Eq.(A.2e), namely, $Z_{ijkl}\sigma_{ij}P_kP_lP_mP_n$ and $W_{ijklmn}\sigma_{ij}\sigma_{kl}P_mP_n$.

Since $Q_c > 0$ for BaTiO$_3$, compressive Vegard strains ($w_s < 0$) suppress $P_s$ and can induce the paraelectric (**PE**) state, while tensile Vegard strains ($w_s > 0$) increase $P_s$ and support the ferroelectric (**FE**) state in the BaTiO$_3$ core. For tensile Vegard strains, $P_s$ decreases with an increase of the core radius as shown for the range of radii (5 – 50) nm used in this study. It is seen from the comparison of the parts (a)-(h) in **Figs. 2 – 3** that $P_s$ is the largest for $R_c = 5$ nm and the smallest for $R_c = 50$ nm. Note that the color scale in **Figs. 2 – 3** ranges from the maximal positive value (red color) to the minimal zero value (dark violet color).

The spontaneous polarization calculated using the 2-4-6 LGD free energy, where $Z_{ijk} = 0$ and $W_{ijk} = 0$, is shown in **Fig. 2(a) – 2(d)**. $P_s$ appears at some critical temperature- and size-dependent strain, denoted as $w_{cr}$, and monotonically increases with an increase in $w_s$ to the 6%. The magnitude of $P_s$, which corresponds to $T = 293$ K and a maximal strain ($w_s$) of 6%, does not exceed 35 µC/cm$^2$ for $R_c = 5$ nm and 30 µC/cm$^2$ for $R_c = 50$ nm. For the case of small radii (5 and 10 nm in **Figs. 2(a,b)**), the magnitude of $P_s$ increases up to (80 - 85) µC/cm2 at temperatures above 440 K (see the red region in the top right corner in the left column of **Fig. 2**), which is much higher than the bulk value for 293 K (26 µC/cm$^2$). Such an increase of $P_s$ is unphysical, because the temperature increase must weaken and eventually destroy the long-range order. The reason of the unphysically large $P_s$ is because the 2-4-6 LGD free energy becomes unstable at temperatures above 440 K, and large tensile Vegard strains shift the instability temperature to lower temperatures; therefore, these large magnitudes of $P_s$ look artificial and disagree with available experimental data.

This strongly suggests that calculating the spontaneous polarization using the 2-4-6 LGD free energy is not practical.

The spontaneous polarization calculated using the 2-4-6-8 LGD free energy, where $Z_{ijk} = 0$ and $W_{ijk} = 0$, is shown in **Fig. 2(e) – 2(h)**. The magnitude of $P_s$, which corresponds to $T = 293$ K and maximal $w_s = 6\%$, does not exceed 30 µC/cm² for $R_c = 5$ nm and 25 µC/cm² for $R_c = 50$ nm. The polarization decreases with a temperature increase (e.g., $P_s$ is smaller than 5 µC/cm² for 500 K, $R_c = 25$ nm, and $w_s = 4\%$), which is physical. The polarization increases up to 30 µC/cm² with the temperature decrease to 100 K (see the red region in the bottom right corner in the right column of **Fig. 2**), which is reasonable because the lowering temperature supports long-range order. Thus, the 2-4-6-8 LGD free energy being stable, can be used for a better description of BaTiO3 core-shell nanoparticles at arbitrary temperatures and high Vegard strains. However, the magnitude of $P_s$ does not exceed 35 µC/cm² which is too small in comparison with the experimentally observed giant values that exceed 120 µC/cm² [6, 11].

Next, we study how the nonlinear electrostriction coupling, under the condition $W_c \geq 0$, influences the spontaneous polarization of the BaTiO3 core-shell nanoparticles. In this case, the nonlinear electrostriction coupling does not negatively impact the stability of the 2-4-6-8 LGD functional. We searched numerically for optimal values of $W_c$ and $Z_c$ corresponding to the maximal spontaneous polarization over the full temperature range of the investigation, (0 - 500) K, and, at the same time, values that are small enough for the validity of elastically-isotropic approximation for $W_{ijk}$ and $Z_{ijk}$ in the core-shell nanoparticle. The obtained optimal values are $W_c = 0$ and $Z_c = 0.28$ m⁸/C⁴ for the 2-4-6 LGD free energy functional, and $W_c = 1.2 \cdot 10^{-12}$ m⁴/(Pa·C²) and $Z_c = 0.44$ m⁸/C⁴ for the 2-4-6-8 LGD free energy functional.

The spontaneous polarization, $P_s$, calculated using the 2-4-6 LGD free energy with nonlinear electrostriction coupling, is shown in **Fig. 3(a) – 3(d)**. $P_s$ appears at some critical temperature- and size-dependent strain, denoted as $w_{cr}$, and monotonically increases with an increase in $w_s$ to the 6%. The magnitude of $P_s$ reaches 55 µC/cm² for $R_c = 5$ nm and 50 µC/cm2 for $R_c = 50$ nm, $T = 293$ K, and $w_s = 6\%$. Note that $P_s$ strongly increases (> 65 µC/cm²) with temperature increase above 440 K only for the 2-4-6 LGD model (the red top right corners are present only in the left column of **Fig. 3**). Thus, the consideration of the coupling between the Vegard strains and nonlinear electrostriction in the 2-4-6 LGD free energy allows $P_s$ to increase more than a factor of two in comparison with the bulk value, however, it does not prevent the functional instability at the temperatures higher than 440 K.

The spontaneous polarization $P_s$ calculated using the 2-4-6-8 LGD free energy with nonlinear electrostriction coupling is shown in **Fig. 3(e) – 3(h)**. For the 2-4-6-8 LGD model, $P_s$ increases with a strain increase and does not increase with temperature. A red color depicts high strains in the right column of **Fig. 3**, and the color saturation increases very weakly as $T$ decreases. Also, $P_s$ decreases with a decrease of $w_s$ and vanishes for $w_s$ close to $w_{cr}$, which is natural and expected. For $w_s > w_{cr}$ the magnitude of $P_s$ increases with an increase in $w_s$ due to the nonlinear electrostriction coupling. The magnitude of $P_s$ reaches 50 µC/cm² for $R_c = 5$ nm and 45 µC/cm² for $R_c = 50$ nm at $T = 293$ K and $w_s = 6\%$. Thus, the consideration of the nonlinear electrostriction coupling in the 2-4-6-8 LGD free energy allows $P_s$ to increase a two factor in comparison with the values calculated for 2-4-6-8 LGD free energy with $Z_{ijk} = 0$ and $W_{ijk} = 0$ shown in **Fig. 2(e) – 2(h)**. However, the maximal values of $P_s$, which do not exceed 55 µC/cm², are more than two times smaller in comparison with the experimentally observed values [6, 11].

For the plots in **Fig. 4**, the 2-4-6-8 LGD free energy functional is used with optimal values of $W_c$ and $Z_c$ (i.e., $W_c = 1.2 \cdot 10^{-12}$ m⁴/(Pa·C²) and $Z_c = 0.44$ m⁸/C⁴). Note the calculations of the strains and tetragonality ratio using 2-4-6 LGD free energy with optimal values of $W_c$ and $Z_c$ (i.e., $W_c = 0$ and $Z_c = 0.27$ m⁸/C⁴) give similar results to those shown in **Fig. 4(a)-4(d)** for the strains $u_1^c$ and $u_3^c$ and in **Fig. 4(e)-4(f)** for $\frac{c}{a}$ only at temperatures below 400 K. Note, the strains and tetragonality ratio calculated using 2-4-6 LGD free energy start to increase for temperatures above 400 K, and, in particular, $\frac{c}{a}$ becomes larger than 1.1 when the temperature increases above 440 K. Since $\frac{c}{a} - 1 \approx u_3^c - u_1^c \approx (Q_{11} - Q_{12})P_s^2$ in accordance with Eq.(7), the increase of $u_1^c, u_3^c$, and $\frac{c}{a}$ for $T>400$ K is a direct consequence of the spontaneous polarization increase for temperatures $T>400$ K (as shown in **Fig. 3(a)-(d)**). Since the unphysical increase of $P_s^2$ happens for $T>400$ K due to the 2-4-6 LGD free energy inapplicability for temperatures above 440 K, one cannot trust the increase of $u_1^c, u_3^c$, and $\frac{c}{a}$ for $T>400$ K, and thus we do not show the figure for the strains and tetragonality ratio calculated using 2-4-6 LGD free energy in this work.

The spontaneous polarization value 50 µC/cm² corresponds to tensile core strains, $u_1^c$ and $u_3^c$, as high as (3 – 6) % (see **Figs. 4(a)-4(d)**) and tetragonality ratios as high as 1.02 – 1.04 (see **Fig. 4(e)-4(f)**), although this magnitude of the spontaneous polarization is nowhere near the large experimentally measured values. A spontaneous polarization larger than 50 µC/cm² can be reached by the application of higher tensile strains and/or high positive intrinsic surface stresses (note, compressive stains do not contribute to an increase in polarization), however, this would lead to unphysically high tetragonality ratios. In particular, a 27% strain difference (i.e., a tetragonality ratio as high as 1.27) would be required to match the experimentally observed polarization value of 130 µC/cm² [6], which seems physically impossible. This means that physical mechanisms, other than the Vegard and/or mismatch strains, linear and nonlinear electrostriction couplings, which are not considered here, are responsible for the polarization enhancement higher than 100 µC/cm².

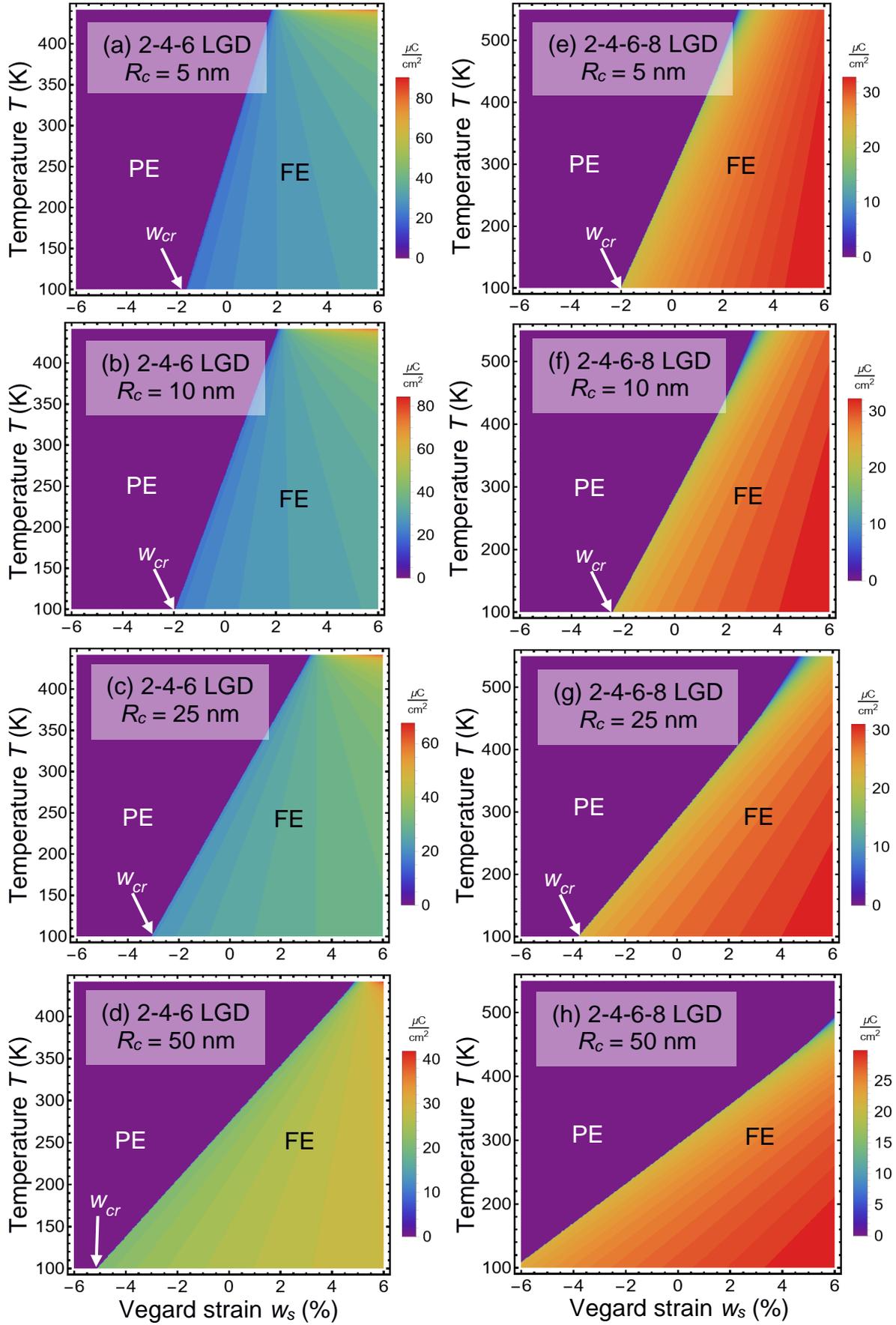

**FIGURE 2.** The spontaneous polarization $P_s$ of a single-domain BaTiO$_3$ core-shell nanoparticle with radius $R_c = 5$ nm (**a, e**), 10 nm (**b, f**), 25 nm (**c, g**), and 50 nm (**d, h**) calculated as a function of temperature $T$ and Vegard strain $w_s$ using the 2-4-6 (**a-d**) and 2-4-6-8 (**e-h**) LGD free energy without nonlinear electrostriction coupling, $\lambda = 0.5$ nm, $\Delta R = 5$ nm,

and $\varepsilon_s = 500$. Color coding in the diagrams is the absolute value of $P_s$ in the deepest potential well of the LGD free energy. The color scale shows the $P_s$ in µC/cm$^2$. The abbreviations "PE" and "FE" refer to the paraelectric and ferroelectric phases, respectively.

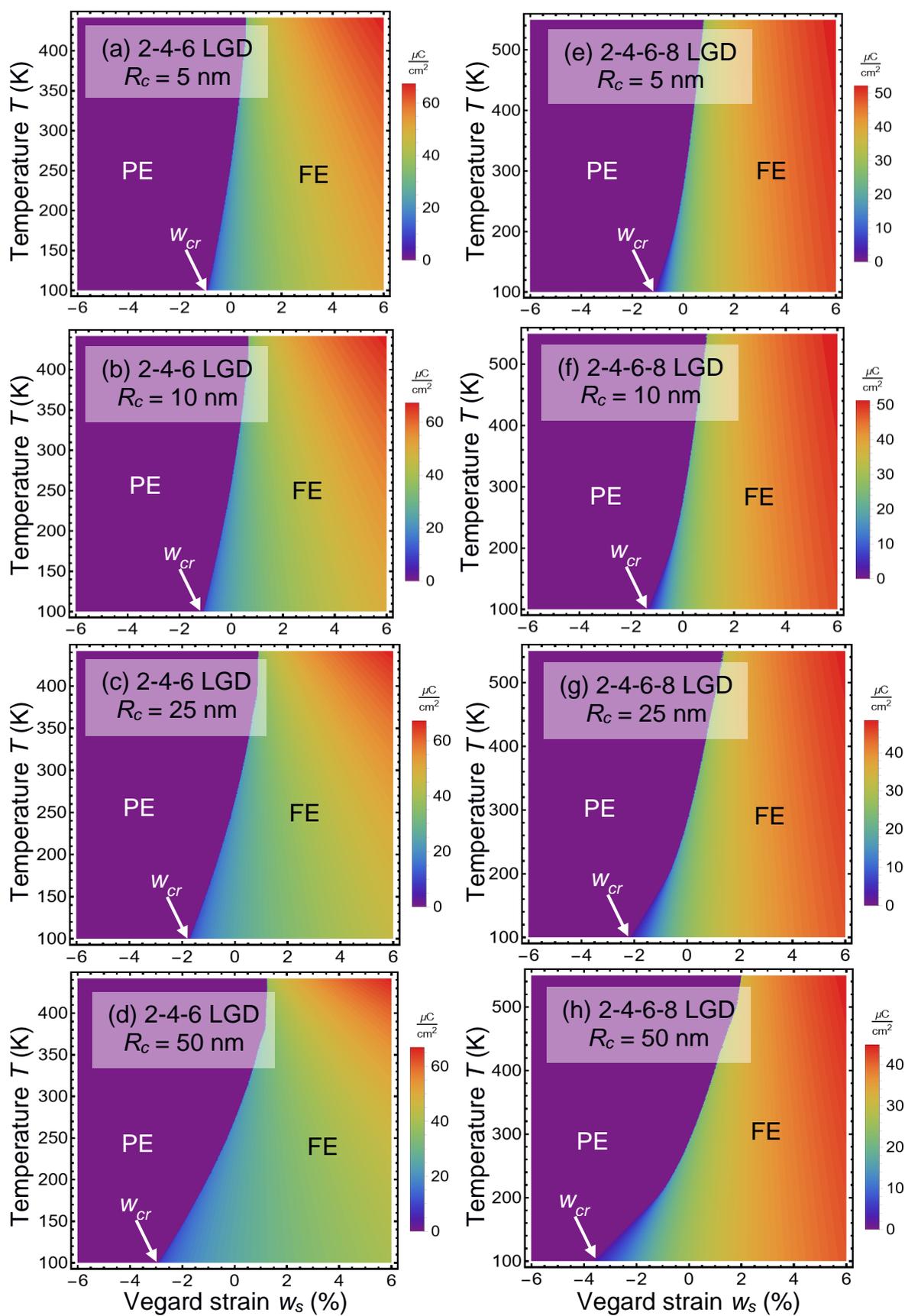

**FIGURE 3.** The spontaneous polarization $P_s$ of a single-domain BaTiO$_3$ core-shell nanoparticle with radius $R_c = 5$ nm

(a, e), 10 nm (b, f), 25 nm (c, g), and 50 nm (d, h) calculated as a function of temperature $T$ and Vegard strain $w_s$ using optimal nonlinear electrostriction values. For 2-4-6 LGD free energy, $W_c = 0$ and $Z_c = 0.27$ m$^8$/C$^4$ (a-d); and 2-4-6-8 LGD free energy, $W_c = 1.2 \cdot 10^{-12}$ m$^4$/(Pa·C$^2$) and $Z_c = 0.44$ m$^8$/C$^4$ (e-h). Other parameters and designations are the same as in **Fig. 2.**

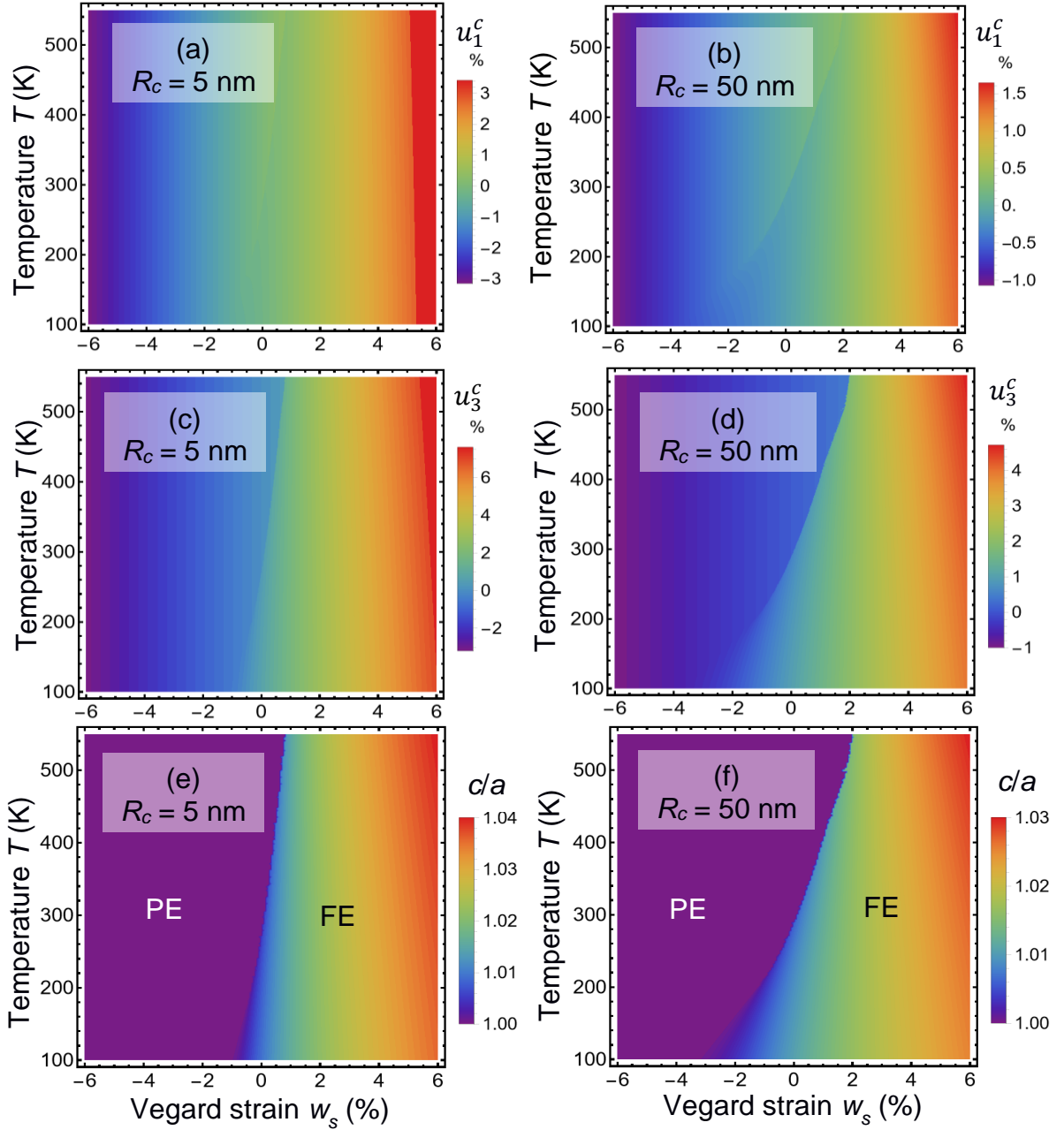

**FIGURE 4.** The strain components $u_1^c$ (a, b) and $u_3^c$ (c, d), and the tetragonality ratio $c/a$ (e, f) of the single-domain BaTiO$_3$ core with radius $R_c = 5$ nm (a, c, e) and 50 nm (b, d, f) calculated as a function of temperature $T$ and Vegard strain $w_s$ using the 2-4-6-8 LGD free energy; $W_c = 1.2 \cdot 10^{-12}$ m$^4$/(Pa·C$^2$) and $Z_c = 0.44$ m$^8$/C$^4$. Other parameters are the same as in **Fig. 2.**

As it is seen from **Figs. 2 – 4**, both 2-4-6 and 2-4-6-8 LGD free energies with or without nonlinear electrostriction coupling give less than a half (~50 µC/cm$^2$) of what is measured for the spontaneous polarization at room temperature (~100 - 130 µC/cm$^2$), yet they have very large tetragonality ratios. It also should be noted that the polarization of core-shell nanoparticles can depend critically on the preparation method: ball-milled nanoparticles reveal $P_S \cong 130$ μC/cm$^2$, while much smaller $P_S$ values are measured in non-milled nanoparticles [6, 11]. The discrepancy of more than a factor of two between LGD models and experiments [6, 11] may be explained by other factors not considered in this work. It could be the dependence of the linear electrostriction coefficients $Q_{ij}$ and elastic modulus $s_{ij}$ on the preparation methods of the nanoparticles or post fabrication techniques. However, recent work reveals an extraordinarily high electrostriction due to interface effects [41]. As a rule, the influence of the interface is important near the surface (i.e., several nm from the surface); but the scale may be greater for milled nanoparticles of (5 – 10) nm radius, which have a quasi-cubic or irregular shape, because of their evolved surfaces and corners that contribute to the formation of inhomogeneous internal strains. Note that the elastic mismatch at the core-shell interface and internal strains can also influence the penetration depth of the surface-induced electrostriction: as a rule, the stronger the mismatch and/or strains, the greater the depth can be. Another possibility may be a negative extrapolation length in the boundary conditions for polarization, which would support the surface-induced polarization enhancement. In this work, we imposed natural boundary conditions, which correspond to the infinite extrapolation length. However, the effect of the extrapolation length can be very "short-range" [44], meaning that the polarization enhancement induced by the negative extrapolation length would be significant only in a thin sub-surface layer, as thin as (3 – 10) lattice constants.

Below we will show that, due to the polarization enhancement, the Vegard strains can improve the electrocaloric properties of a core-shell ferroelectric nanoparticle, which can be important for applications such as nanocoolers. The relative electrocaloric (**EC**) temperature change, $\Delta T_{EC} = T_{FE} - T$, can be calculated from the expression [45]:

$$\Delta T_{EC} = -T \int_{E_1}^{E_2} \frac{1}{\rho C_P} \left(\frac{\partial P_3}{\partial T}\right)_E dE \approx \frac{T}{\rho C_P} \left(\frac{\alpha_T}{2}[P_3^2(E_2) - P_3^2(E_1)] + \frac{\beta_T}{4}[P_3^4(E_2) - P_3^4(E_1)] + \frac{\gamma_T}{6}[P_3^6(E_2) - P_3^6(E_1)]\right), \quad (8a)$$

where $\rho_P$ is the volume density and $C_P$ is the specific heat of the nanoparticle core; $T$ is the ambient temperature, $T_{FE}$ is the temperature of the ferroelectric core measured in adiabatic conditions; $E_1$ and $E_2$ are the values of the quasi-static electric field $E_3^e$ applied to the nanoparticle in adiabatic conditions; and coefficients $\alpha_T = \frac{\partial \alpha_R}{\partial T}$, $\beta_T = \frac{\partial \beta_R}{\partial T}$, and $\gamma_T = \frac{\partial \gamma_R}{\partial T}$ ($\alpha_R$, $\beta_R$, and $\gamma_R$ are introduced in Eqs.(6)).

Note, that we are especially interested in reaching a maximal negative $\Delta T_{EC} < 0$ corresponding to EC cooling of the nanoparticle, which is required for nanocooler-based applications. To reach the maximal $\Delta T_{EC} < 0$, it is necessary to set $E_1 = 0$ and $E_2 = E_c$ in Eq.(8a), where $E_c$ is the coercive field of the core-shell nanoparticle, and determine that the greater polarization corresponds to the larger absolute value of $\Delta T_{EC}$:

$$\Delta T_{EC} \approx -\frac{T}{\rho C_P}\left(\frac{\alpha_T}{2}P_S^2 + \frac{\beta_T}{4}P_S^4 + \frac{\gamma_T}{6}P_S^6\right). \quad (8b)$$

In Eq.(8b), $P_3^2(0) = P_S^2$ and $P_3^2(E_c) = 0$. Since $\alpha_T > 0$, $\beta_T > 0$, and $\gamma_T < 0$ for BaTiO$_3$, the EC effect is negative (cooling, $\Delta T_{EC} < 0$) for small and intermediate values of $P_S^2$; however, it can change the sign (heating, $\Delta T_{EC} > 0$) for very large values of the spontaneous polarization.

The specific heat depends on polarization for ferroelectrics and can be modeled as following [46]:

$$C_P = C_P^0 - T\frac{\partial^2 g}{\partial T^2}, \quad (8c)$$

where $C_P^0$ is the polarization-independent part of specific heat and $g$ is the density of the LGD free energy (A.2). According to experimental results, the specific heat usually has a maximum at the point (i.e., coordinates of temperature and radius/strain) of the first order ferroelectric phase transition, and the maximum height is about (10 – 30) % of the $C_p$ value near $T_C$ (see e.g., Ref. [47]). For estimates of the EC temperature change, we assume that the mass density and heat-capacitance of the BaTiO$_3$ are $\rho_P = 6.02 \cdot 10^3$ kg/m$^3$ and $C_P = 4.6 \cdot 10^2$ J/(kg K), respectively.

The calculations of the EC response are performed over a wide temperature range, (250 – 400) K; for a range of core radii, $5 < R_c < 50$ nm, for which the Vegard strains are pronounced. The nanoparticle core must be in the ferroelectric state to produce any noticeable EC cooling (e.g., $\Delta T_{EC} < -2$ K); when the core is in the paraelectric state, only a weak EC heating (e.g., $0 < \Delta T_{EC} < 2$ K) is possible [45]. The spontaneous polarization, $P_s$, and the electrocaloric temperature change, $\Delta T_{EC}$, as a function of the core radius $R_c$ and Vegard strain $w_s$, are shown in **Fig. 5**. $P_s$ is calculated for $E_3^e \to 0$, and $\Delta T_{EC}$ is calculated for $E_3^e \to E_c$, where $E_c$ is the coercive field. Since $Q_c > 0$, compressive Vegard strains $w_s < 0$ suppress $P_s$ and can induce the PE state, and tensile Vegard strains $w_s > 0$ increase $P_s$ and support the FE state in the BaTiO$_3$ core. Due to the synergy of the size effect and tensile Vegard strain, the spontaneous polarization reaches (50 – 30) µC/cm$^2$ in the temperature range (298 – 388) K, which is higher than the corresponding values (26 – 21) µC/cm$^2$ for a bulk BaTiO$_3$. For the same temperature range, the EC cooling of the BaTiO$_3$ core can be more than 6 K far from the boundary of the size-induced paraelectric-ferroelectric phase transition. Note that the EC cooling cannot exceed a temperature change of 2.5 K for a bulk BaTiO$_3$ nor for unstrained BaTiO$_3$ nanoparticles in accordance with earlier LGD-based theoretical calculations [45]. The tensile Veard strain increases the spontaneous polarization, which in turn increases the EC temperature change according to Eq.(8b).

The change in the sign of the EC effect with an increase in the spontaneous polarization is related to the features of the LGD free energy for BaTiO$_3$, where not only the coefficient at the 2-nd power of polarization depends on temperature, but also the coefficients at the 4-th and 6-th powers. According to Eq.(8a), this leads to several contributions to the electrocaloric effect, which are proportional to the corresponding polarization powers. Since the coefficient $\gamma_T$ is negative, the EC effect changes sign for large values of the polarization.

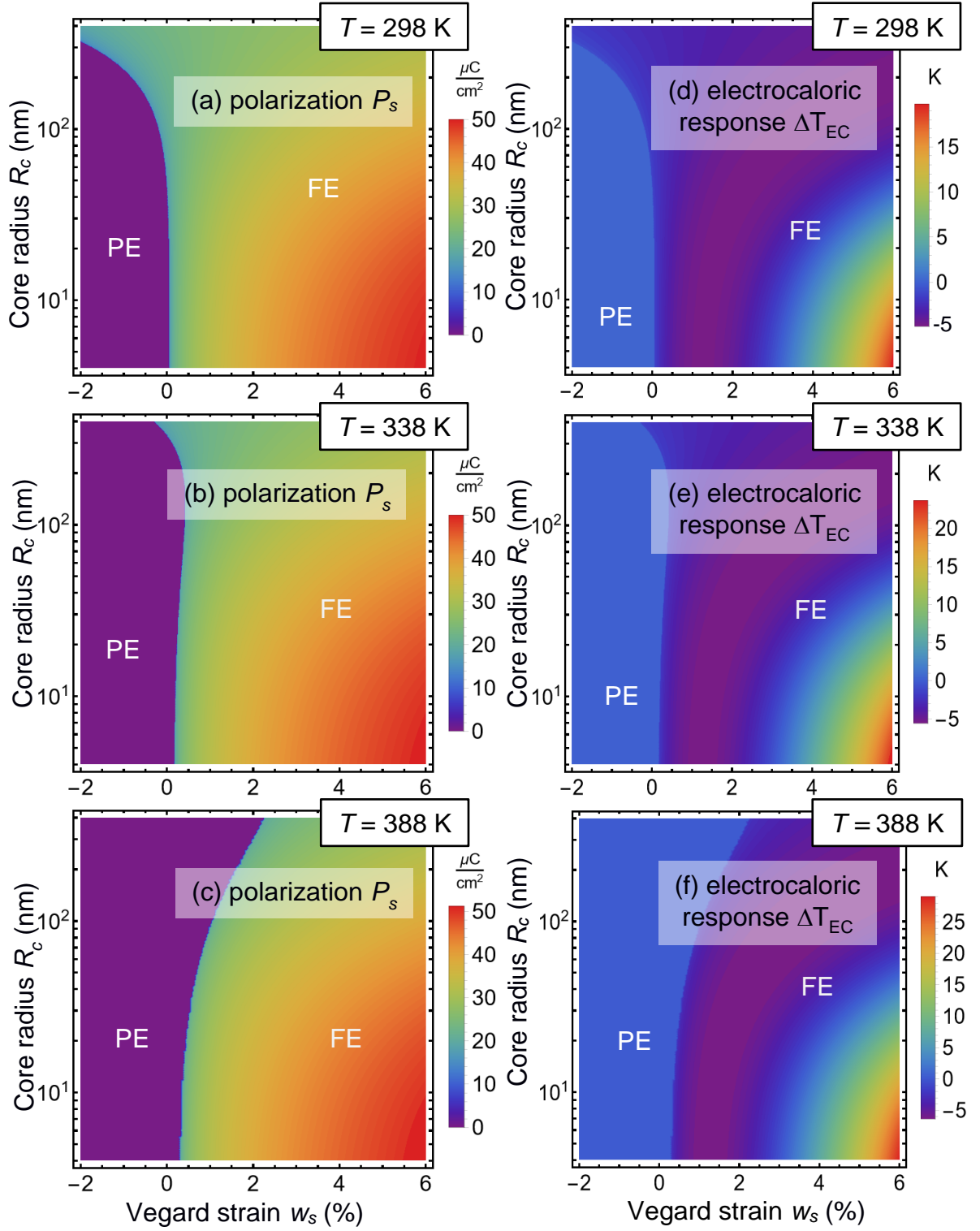

**FIGURE 5.** The spontaneous polarization $P_s$ (**a, b, c**) and the electrocaloric temperature change $\Delta T_{EC}$ (**d, e, f**) of a single-domain BaTiO$_3$ core-shell nanoparticle calculated as a function of core radius $R_c$ and Vegard strain $w_s$ using the 2-4-6-8 LGD free energy, $W_c = 1.2 \cdot 10^{-12}$ m$^4$/(Pa·C$^2$) and $Z_c = 0.44$ C$^{-4}$·m$^8$, and temperatures $T =$298 K **(a, d)**, 338 K **(b, e)**, and 388 K **(c, f)**. Color scale corresponds to $P_S$ in μC/cm$^2$ and $\Delta T_{EC}$ in K. Other parameters are the same as in **Fig. 2.**

## IV. CONCLUSIONS

This work provides a systematic analytical description of the polar properties of core-shell BaTiO$_3$ nanoparticles using the 2-4-6-8 Landau-Ginzburg-Devonshire free energy functional, which considers nonlinear electrostriction coupling and large Vegard strains in the shell. We revealed that the spontaneous polarization, as high as 50 μC/cm$^2$, can be stable in the BaTiO$_3$ core with radius 5 - 50 nm at room temperature, if a 5-nm paraelectric shell is stretched (3-6)% by the Vegard strains. We can conclude that the nonlinear electrostriction coupling in the core and tensile Vegard strains in the shell are key physical factors of a spontaneous polarization enhancement.

The polarization 50 μC/cm$^2$ corresponds to tetragonality ratios as high as (1.02 – 1.04). The application of higher strains and/or surface stresses would lead to unphysically high tetragonality ratios. In particular, the experimentally observed polarization 130 μC/cm2 [6] corresponds to tetragonality ratios as high as 1.27.

The value ~50 μC/cm$^2$ is less than a half of what is measured for the spontaneous polarization at room temperature (~130 μC/cm$^2$) for ball-milled core-shell BaTiO$_3$ nanoparticles. A discrepancy of more than a factor of two between the considered 2-4-6 and 2-4-6-8 LGD models and experiments [6, 11] may be explained by several factors not considered in this work. It can be the dependence of the anisotropic linear electrostriction coefficients $Q_{ij}$ on the preparation way of the nanoparticles, which can be extraordinarily high due to the interface effects [41]. $Q_{ij}$ can reach giant values for milled nanoparticles of (5 – 10) nm radius and quasi-cubic shape, because evolved surface and corners contribute to the formation of inhomogeneous internal strains. Another possibility is a negative extrapolation length in the boundary conditions for polarization, which would support the surface-induced polarization enhancement, despite the extrapolation length effect is short-range [44].

The Vegard strains can improve the electrocaloric properties of a core-shell ferroelectric nanoparticle due to the strain-induced polarization enhancement. In particular, the tensile Vegard strain of (3-6)% increase the spontaneous polarization up to 50 μC/cm$^2$, and the spontaneous polarization in turn lowering the EC cooling temperature up to 6 K (in comparison with an unstrained bulk BaTiO$_3$, where the change in the EC temperature cannot exceed 2.5 K). Thus, the dense nanocomposites containing core-shell BaTiO$_3$ nanoparticles can be important for applications as nanocoolers.

**Acknowledgments.** A.N.M. acknowledges EOARD project 9IOE063 and related STCU partner project P751a. E.A.E. acknowledges the National Academy of Sciences of Ukraine. S.V.K. is supported by the center for 3D Ferroelectric Microelectronics (3DFeM), an Energy Frontier Research Center funded by the U.S. Department of Energy (DOE), Office of Science, Basic Energy Sciences under Award Number DE-SC0021118.

**Authors contribution.** E.A.E. wrote the codes, prepare figures, and estimated the nonlinear electrostriction couplings. A.N.M. generated the research idea and formulated the electrostatic and elastic problems, performed most of the analytical calculations and wrote the paper draft. S.V.K. and D.R.E. worked on the results interpretation, discussion, and paper improvement.

**APPENDIX A. Mathematical formulation of the problem and computation details**

**A1. Electric field and LGD free energy of the core-shell nanoparticle**

We consider a ferroelectric nanoparticle core of radius $R_c$ with a three-component ferroelectric polarization vector $\boldsymbol{P}$. The core is regarded as insulating, without any free charges. It is covered with a semiconducting paraelectric shell of thickness $\Delta R$ that is characterized by an isotropic relative dielectric permittivity tensor $\varepsilon_{ij}^S = \delta_{ij}\varepsilon_s$. The core-shell nanoparticle is placed in a dielectric medium (polymer, gas, liquid, air, or vacuum) with an effective dielectric permittivity, $\varepsilon_e$. The core-shell geometry is shown in **Fig. 1** of the main text.

Since the ferroelectric polarization contains background and soft mode contributions, the electric displacement vector has the form $\boldsymbol{D} = \varepsilon_0\varepsilon_b\boldsymbol{E} + \boldsymbol{P}$ inside the core. In this expression $\varepsilon_b$ is a relative background permittivity of the core [48], $\varepsilon_0$ is the universal dielectric constant, and $\boldsymbol{P}$ is a ferroelectric polarization containing the spontaneous and field-induced contributions. As a rule, $4 < \varepsilon_b < 10$. The expression $D_i = \varepsilon_0\varepsilon_{ij}^S E_j$ is valid in the shell and $D_i = \varepsilon_0\varepsilon_e E_i$ in is valid in the isotropic effective medium.

The electric field components, $E_i$, are derived from the electric potential φ in a conventional way, $E_i = -\partial\varphi/\partial x_i$. The potential $\varphi_f$ satisfies the Poisson equation in the ferroelectric core (subscript "*f*"):

$$\varepsilon_0\varepsilon_b \left(\frac{\partial^2}{\partial x_1^2} + \frac{\partial^2}{\partial x_2^2} + \frac{\partial^2}{\partial x_3^2}\right) \varphi_f = \frac{\partial P_i}{\partial x_i}, \qquad 0 \leq r \leq R_c, \tag{A.1a}$$

and a Debye-type equation in the shell (subscript "*s*"):

$$\left(\frac{\partial^2}{\partial x_1^2} + \frac{\partial^2}{\partial x_2^2} + \frac{\partial^2}{\partial x_3^2}\right) \varphi_s = -\frac{\varphi_s}{\lambda^2}, \qquad R_c < r \leq R_s, \tag{A.1b}$$

where $\lambda = \sqrt{\frac{\varepsilon_0\varepsilon_s k_B T}{e^2 n}}$ is the Debye screening length defined by the concentration of free carriers $n$ in the shell.

The electric potential $\varphi_e$ in the external region outside the shell satisfies the Laplace equation (subscript "*e*"):

$$\varepsilon_0\varepsilon_e \left(\frac{\partial^2}{\partial x_1^2} + \frac{\partial^2}{\partial x_2^2} + \frac{\partial^2}{\partial x_3^2}\right) \varphi_e = 0, \qquad r > R_s. \tag{A.1c}$$

Equations (A.1) should be solved with the continuity conditions for the electric potential and normal components of the electric displacements at the particle surface and core-shell interface:

$$(\varphi_e - \varphi_s)|_{r=R_s} = 0, \quad \boldsymbol{n}(\boldsymbol{D}_e - \boldsymbol{D}_s)|_{r=R_s} = 0, \tag{A.1d}$$

$$(\varphi_s - \varphi_f)|_{r=R_c} = 0, \quad \boldsymbol{n}(\boldsymbol{D}_s - \boldsymbol{D}_f)|_{r=R_c} = 0. \tag{A.1e}$$

Either charges are absent or the applied voltage is fixed at the boundaries of the computation region:

$$\left.\frac{\partial \varphi_e}{\partial x_l} n_l\right|_{x=\pm\frac{L}{2}} = 0, \quad \left.\frac{\partial \varphi_e}{\partial x_l} n_l\right|_{y=\pm\frac{L}{2}} = 0, \quad \varphi_e|_{z=+\frac{L}{2}} = 0, \quad \varphi_e|_{z=-\frac{L}{2}} = V_e, \tag{A.1f}$$

where $V_e$ is the applied voltage difference and $L$ is the size of computation region.

The LGD free energy functional $G$ of the core polarization $\boldsymbol{P}$ additively includes a Landau expansion on 2-nd, 4-th, 6-th, and 8-th powers of the polarization, $G_{Landau}$; a polarization gradient energy contribution, $G_{grad}$; an electrostatic contribution, $G_{el}$; the elastic, linear, and nonlinear electrostriction couplings and flexoelectric contributions, $G_{es+flexo}$; and a surface energy, $G_S$. The functional $G$ has the form [49, 50, 51]:

$$G = G_{Landau} + G_{grad} + G_{el} + G_{es+flexo} + G_{VS} + G_S, \tag{A.2a}$$

$$G_{Landau} = \int_{0<r<R_c} d^3r \left[a_i P_i^2 + a_{ij} P_i^2 P_j^2 + a_{ijk} P_i^2 P_j^2 P_k^2 + a_{ijkl} P_i^2 P_j^2 P_k^2 P_l^2\right], \tag{A.2b}$$

$$G_{grad} = \int_{0<r<R} d^3r \frac{g_{ijkl}}{2} \frac{\partial P_i}{\partial x_j} \frac{\partial P_k}{\partial x_l}, \tag{A.2c}$$

$$G_{el} = -\int_{0<r<R_c} d^3r \left(P_i E_i + \frac{\varepsilon_0 \varepsilon_b}{2} E_i E_i\right) - \frac{\varepsilon_0}{2} \int_{R_c<r<R_s} \varepsilon_{ij}^S E_i E_j d^3r - \frac{\varepsilon_0}{2} \int_{r>R+\Delta R} \varepsilon_{ij}^e E_i E_j d^3r, \tag{A.2d}$$

$$G_{es+flexo} = -\int_{0<r<R_c} d^3r \left(\frac{s_{ijkl}}{2} \sigma_{ij} \sigma_{kl} + Q_{ijkl} \sigma_{ij} P_k P_l + Z_{ijkl} \sigma_{ij} P_k P_l P_m P_n + \frac{1}{2} W_{ijklmn} \sigma_{ij} \sigma_{kl} P_m P_n + F_{ijkl} \sigma_{ij} \frac{\partial P_l}{\partial x_k}\right) \tag{A.2e}$$

$$G_S = \frac{1}{2} \int_{r=R_c} d^2r \, a_{ij}^{(S)} P_i P_j. \tag{A.2g}$$

The coefficient $a_i$ linearly depends on temperature $T$:

$$a_i(T) = \alpha_T [T - T_C(R_c)], \tag{A.3a}$$

where $\alpha_T$ is the inverse Curie-Weiss constant and $T_C(R_c)$ is the ferroelectric Curie temperature renormalized by electrostriction and surface tension as [49, 50]:

$$T_C(R_c) = T_C \left(1 - \frac{Q_c}{\alpha_T T_C} \frac{2\mu}{R_c}\right), \tag{A.3b}$$

where $T_C$ is a Curie temperature of a bulk ferroelectric. $Q_c$ is the sum of the electrostriction tensor diagonal components, which is positive for most ferroelectric perovskites with cubic m3m symmetry in the paraelectric phase, namely $0.004 < Q_c < 0.04 \text{m}^4/\text{C}^2$.

Tensor components $a_{ij}$, $a_{ijk}$, and $a_{ijkl}$ are listed in **Table AI.** The gradient coefficients tensor $g_{ijkl}$ are positively defined and regarded as temperature-independent. The following designations are used in Eq.(A.2e): $\sigma_{ij}$ is the stress tensor, $s_{ijkl}$ is the elastic compliances tensor, $Q_{ijkl}$, $Z_{ijklmn}$, and $W_{ijklmn}$ are the linear and two nonlinear electrostriction tensors, whose values and/or ranges are listed in **Table AI**.

Since µ is relatively small, not more than (1 – 4) N/m for most perovskites, and to focus on the influence of linear and nonlinear electrostriction effects, we do not consider the surface tension and flexoelectric coupling in this work and set $\mu = 0$ and $F_{ijkl} = 0$.

Allowing for the Khalatnikov mechanism of polarization relaxation [52], minimization of the free energy (A.2) with respect to polarization leads to three coupled time-dependent Euler-Lagrange equations for polarization components inside the core,

$$\frac{\delta G}{\delta P_i} = -\Gamma \frac{\partial P_i}{\partial t}, \tag{A.4}$$

where the subscript $i = 1, 2, 3$, and $\Gamma$ and is the temperature-dependent Khalatnikov coefficient [53].

The boundary condition for polarization at the core-shell interface $r = R_c$ is:

$$a_{ij}^{(S)} P_j + g_{ijkl} \frac{\partial P_k}{\partial x_l} n_j \bigg|_{r=R_c} = 0, \tag{A.5}$$

where $n_j$ are the components of the outer normal to the surface, and the subscripts $\{i, j, k, l\} = \{1, 2, 3\}$. Below we use the so-called "natural" boundary conditions corresponding to $a_{ij}^{(S)} = 0$, which support the formation of a single-domain state in the core.

Elastic stresses satisfy the equation of mechanical equilibrium in the computation region,

$$\frac{\partial \sigma_{ij}}{\partial x_j} = 0, \qquad -L/2 < \{x, y, z\} < L/2. \tag{A.6}$$

Elastic equations of state follow from the variation of the energy (A.2e) with respect to elastic stress, $\frac{\delta G}{\delta \sigma_{ij}} = -u_{ij}$, namely:

$$s_{ijkl}\sigma_{kl} + Q_{ijkl}P_k P_l + Z_{ijkl}P_k P_l P_m P_n + W_{ijklmn}\sigma_{kl}P_m P_n = u_{ij}, \tag{A.7}$$

where $0 < r \leq R_c$, $u_{ij}$ is the strain tensor components related to the displacement components $U_i$ in the following way: $u_{ij} = (\partial U_i/\partial x_j + \partial U_j/\partial x_i)/2$.

The elastic displacement components $U_i$ and normal stresses $\sigma_{ij}$ are continuous functions at the core-shell interface ($r = R_c$):

$$U_i|_{r=R_c-0} = U_i|_{r=R_c+0}, \qquad \sigma_{ij}n_j\big|_{r=R_c-0} = \sigma_{ik}n_k\big|_{r=R_c+0}, \tag{A.8a}$$

as well as at the interface between the shell and the external media ($r = R_s$):

$$U_i|_{r=R_s-0} = U_i|_{r=R_s+0}, \qquad \sigma_{ij}n_j\big|_{r=R_s-0} = \sigma_{ik}n_k\big|_{r=R_s+0}. \tag{A.8b}$$

**Table AI.** LGD coefficients and other material parameters of a BaTiO$_3$ core in Voigt notations

| Parameter, its description, and dimension (in the brackets) | Numerical value or variation range | |
|---|---|---|
| | **2-4-6 LGD expansion** | **2-4-6-8 LGD expansion** |
| Expansion coefficients $a_i$ in the term $a_i P_i^2$ in Eq.(A.2b) (C$^{-2}$·mJ) | $a_1 = 3.34 \times 10^5 (T-381)$ <br> $\alpha_T = 3.34 \times 10^5$ <br> ($a_1 = -2.94 \times 10^7$ at 298 K) | $a_1 = 3.61(T-391) \times 10^5$ |
| Expansion coefficients $a_{ij}$ in the term | $a_{11} = 4.69(T-393) \times 10^6 - 2.02 \times 10^8$ <br> $a_{12} = 3.230 \times 10^8$ | $a_{11} = 4.0 \times 10^6 T - 1.83 \times 10^9$ <br> $a_{12} = 6.7 \times 10^6 T - 2.24 \times 10^9$ |
| $a_{ij}P_i^2 P_j^2$ in Eq.(A.2b) (C$^{-4}$·m$^5$J) | ($a_{11} = -6.71 \times 10^8$ at 298 K) | |
| Expansion coefficients $a_{ijk}$ in the term $a_{ijk}P_i^2 P_j^2 P_k^2$ in Eq.(A.2b) (C$^{-6}$·m$^9$J) | $a_{111} = -5.52(T-393) \times 10^7 + 2.76 \times 10^9$<br>$a_{112} = 4.47 \times 10^9$<br>$a_{123} = 4.91 \times 10^9$<br>($a_{111} = 82.8 \times 10^8$ at 298 K) | $a_{111} = -3.2 \times 10^7 \, T + 1.39 \times 10^{10}$<br>$a_{112} = -2.2 \times 10^9$<br>$a_{123} = 5.51 \times 10^{10}$ |
| Expansion coefficients $a_{ijkl}$ in the term $a_{ijkl}P_i^2 P_j^2 P_k^2 P_l^2$ in Eq.(A.2b) (C$^{-8}$·m$^{13}$J) | $a_{ijkl} = 0$ | $a_{1111} = 4.84 \times 10^{10}$<br>$a_{1112} = 2.53 \times 10^{11}$<br>$a_{1122} = 2.80 \times 10^{11}$<br>$a_{123} = 9.35 \times 10^{10}$ |
| Linear electrostriction tensor $Q_{ijkl}$ in the term $Q_{ijkl}\sigma_{ij}P_k P_l$ in Eq.(A.2e) (C$^{-2}$·m$^4$) | In Voigt notations $Q_{ijkl} \to Q_{ij}$, which are equal to<br>$Q_{11} = 0.11$<br>$Q_{12} = -0.043$<br>$Q_{44} = 0.059$<br>For a spherical nanoparticle the combinations and functions of $Q_{ij}$ are used: $Q_c = \frac{Q_{11}+2Q_{12}}{3} = 0.042$, $Q_c^2 \equiv (Q_c)^2 = 0.001764$ | |
| Nonlinear electrostriction tensor $Z_{ijklmn}$ in the term $Z_{ijklmn}\sigma_{ij}P_k P_l P_m P_n$ in Eq.(A.2e) (C$^{-4}$·m$^8$) | In Voigt notations $Z_{ijklmn} \to Z_{ijk}$ (also $Z_{ijk}^c \equiv Z_{ijk}$). For a spherical nanoparticle the combinations and functions of $Z_{ijk}$ are used: $Z_c = \frac{Z_{111}+2Z_{211}}{3}$, $Z_c^2 \equiv (Z_c)^2$, where $Z_c$ varies in the range $-1 \leq Z_c \leq 1$ as free parameter | |
| | The optimal value $Z_c = 0.28$ | The optimal value $Z_c = 0.44$ |
| Nonlinear electrostriction tensor $W_{ijklmn}$ in the term $W_{ijklmn}\,\sigma_{ij}\sigma_{kl}P_m P_n$ in Eq.(A.2e) (C$^{-2}$·m$^4$ Pa$^{-1}$) | In Voigt notations $W_{ijklmn} \to W_{ijk}$ (also $W_{ijk}^c \equiv W_{ijk}$). For a spherical nanoparticle the combinations and functions of $W_{ijk}$ are used: $W_c = \frac{W_{111}+2W_{112}+2W_{123}+4W_{122}}{3}$, $W_c^2 \equiv (W_c)^2$, where $W_c$ varies in the range $0 \leq W_c \leq 10^{-12}$ as free parameter | |
| | The optimal value $W_c = 0$ | The optimal value $W_c = 1.2 \times 10^{-12}$ |
| Elastic compliances tensor, $s_{ijkl}$, in Eq.(A.2e) (Pa$^{-1}$) | In Voigt notations $s_{ijkl} \to s_{ij}$, also $s_{ij}^c = s_{ij}$, which are equal to<br>$s_{11}^c = s_{11} = 8.3 \times 10^{-12}$<br>$s_{12}^c = s_{12} = -2.7 \times 10^{-12}$<br>$s_{44}^c = s_{44} = 9.24 \times 10^{-12}$ | |
| Polarization gradient coefficients $g_{ijkl}$ in Eq.(A.2c) (C$^{-2}$m$^3$J) | In Voigt notations $g_{ijkl} \to g_{ij}$, which are equal to:<br>$g_{11} = 1.0 \times 10^{-10}$<br>$g_{12} = 0.3 \times 10^{-10}$<br>$g_{44} = 0.2 \times 10^{-10}$ | |
| Surface energy coefficients $a_{ij}^{(S)}$ in Eq.(A.2f) | 0<br>(that corresponds to the natural boundary conditions) | |
| Core radius $R_c$ (nm) | Variable: 5 - 50 | |

| Parameter, its description and dimension (in the brackets) | Numerical value or variation range |
|---|---|
| Background permittivity $\varepsilon_b$ in Eq.(A.2d) (unity) | 7 |

\* $\alpha = 2a_1, \beta = 4a_{11}, \gamma = 6a_{111},$ and $\delta = 8a_{1111}$

**Table AII.** Electric and elastic parameters of the shell and core-shell interface

| Parameter, its description and dimension (in the brackets) | Numerical value or variation range |
|---|---|
| Relative permittivity of the shell $\varepsilon_{ij}^s = \delta_{ij}\varepsilon_s$, where $\delta_{ij}$ is the Kronecker-delta symbol (unity) | $\varepsilon_s$ varies in the range $10^2 - 10^3$, we put $\varepsilon_s = 500$ |
| Screening length $\lambda$ (nm) | $\lambda$ varies in the range $(0.1 - 1)$; we put $\lambda = 0.5$ nm |
| Linear electrostriction tensor $Q_{ijkl}^s$ ($C^{-2} \cdot m^4$) | In Voigt notations $Q_{ijkl}^s \to Q_{ij}^s$, which are close to the core electrostriction $Q_{ij}$ and equal to $Q_{11}^s = 0.11$ $Q_{12}^s = -0.043$ $Q_{44}^s = 0.059$ $Q_s = \frac{Q_{11}^s + 2Q_{12}^s}{3} = 0.042$ |
| Elastic compliances tensor $s_{ij}^s$ (Pa$^{-1}$) | $s_{11}^s = 3.3 \times 10^{-12}$ $s_{12}^s = -0.8 \times 10^{-12}$ $s_{44}^s = 4.1 \times 10^{-12}$ \* |
| Vegard strains in the shell $w_{ij}^s = \delta_{ij} w_s$, where $\delta_{ij}$ is the Kronecker-delta symbol (unity) | The magnitude $w_c$ varies in the range $-0.1 \le w_c \le 0.1$ as free parameter (or from -10 % to +10%) |
| Surface tension coefficient $\mu$ (N/m) | Should be positive and less than $(1-4)$ N/m, we put $\mu = 0$ for the sake of simplicity |
| Shell thickness $\Delta R = R_s - R_c$ (nm) | $0.5 - 5$, we put $\Delta R = 5$ nm |

\* The values of $s_{ij}^s$ for the crystalline shell are smaller than the $s_{ij}^c$ for the crystalline core, which is elastically harder than the BaTiO$_3$ core. An amorphous shell, which is softer than the BaTiO$_3$ core, would correspond to a value of $s_{ij}^s$ larger than the $s_{ij}^c$.

Different sets of electrostriction coefficients $Q_{ij}$ and $Z_{ijk}$ determined from different experiments and/or ab initio calculations are listed in **Table AIII**. In accordance with the table, the value of $Z_c$ can vary from -14.2 m$^8$/C$^4$ to +0.38 m$^8$/C$^4$ in a bulk BaTiO$_3$; and the "scattering" range is a very wide. Since recent works reveal the extraordinarily high electrostriction due to the interface effects [41], we can assume that the possible range of $Z_c$ variation can be even wider in the core-shell BaTiO$_3$ nanoparticles. Indeed, the linear and nonlinear electrostriction coefficients, $Q_{ij}$ and $Z_{ijk}$, should depend on the preparation way of the nanoparticles and their chemical purity. These speculations give us some grounds to vary $Z_c$ within the range from -1.5 m$^8$/C$^4$ to +1.5 m$^8$/C$^4$ looking for the optimal values, which correspond to the highest spontaneous polarization and the best related properties.

**Table AIII.** Different sets of electrostriction coefficients

| $Q_{ij}$ (m$^4$/C$^2$) | $Z_{ijk}$ (m$^8$/C$^4$) | Brief information about the determination way and references |
|---|---|---|
| $Q_c = 0.016$ | 0 | Determined from the dependence of Curie temperature on hydrostatic pressure from the experiment [54] in the supposition $Z_{ijk} = 0$ |
| $Q_{11} = 0.11$, $Q_{12} = -0.045$ $Q_c = 0.020$ | 0 | Determined from the piezoelectric coefficients from the experiment [55] in the supposition $Z_{ijk} = 0$ |
| $Q_c = (0.014 - 0.016)$ | $Z_c = (0.27 - 0.26)$ | Determined from the tricritical point position from the experiments [56, 57, 58] |
| $Q_{11} = 0.118$ $Q_{12} = -0.036$ $Q_c = 0.020$ | $Z_{111} = -0.08$ $Z_{122} = -0.07$ $Z_c = -0.15$ | Determined from the ab initio calculations [59] |
| $Q_{11} = 0.110$ $Q_{12} = -0.045$ $Q_c = 0.020$ | $Z_c = 0.013$ | Determined from the fitting of the transition temperatures dependence on pressure from the experiment [60] |
| $Q_{11} = 0.069$ | $Z_{111} = -14.2$ | Determined from the dependence of Curie temperature on axial pressure from the experiment [61] |
| $Q_{11} = 0.0903$ | 0 | Determined from the dependence of Curie temperature on axial pressure from the experiment [62] |
| $Q_{11} = 0.079$ $Q_{12} = -0.033$ $Q_c = 0.013$ | $Z_{111} = 0.63$ $Z_{122} = -0.52$ $Z_c = -0.41$ | Determined from dependences of Curie temperature on axial and hydrostatic pressure, as well as from the spontaneous strains value using the experiments [63, 64, 62-63] |
| $Q_{11} = 0.057$ $Q_{12} = -0.021$ $Q_c = 0.016$ | $Z_{111} = 0.71$ $Z_{122} = -0.26$ $Z_c = 0.19$ | Determined from the piezoelectric coefficients and spontaneous strains value at room temperature using the experiment [65] |
| $Q_{11} = 0.060$ $Q_{12} = -0.0355$ $Q_c = -0.011$ | $Z_{111} = 0.68$ $Z_{122} = -0.15$ $Z_c = 0.38$ | Determined from the piezoelectric coefficients and spontaneous strains value at room temperature using the experiment [66] |

**A2. The core stress induced by the Vegard strains in the shell**

During the ball-milling the mechanochemical reaction at BaTiO$_3$ nanoparticle surface results in the formation of core-shell nanoparticle. Due to the strong Vegard strains in the shell the elastic mismatch between the core and shell lattices appears and results in the core stress. Below we calculate the stress induced by the Vegard strains in the paraelectric core. We also note that the ball milling makes the particle surface rough, however for the sake of simplicity we assume the particle to be spherical.

In order to find the elastic fields analytically, we use a perturbation approach. At first let us consider an isotropic elastic problem, which has a spherical symmetry, being also consistent with the cubic symmetry of the paraelectric core-shell nanoparticle placed in a soft matter matrix. The elastic displacement in a spherical coordinate frame is given by expression, $\mathbf{U} = \{U_r(r), U_\theta(r), U_\phi(r)\}$, where $U_\theta = U_\phi = 0$ for a spherically-symmetric case. In this case, the displacement vector satisfies the equation [67]

$$\text{grad}(div\mathbf{U}) \equiv \frac{\partial}{\partial r}\frac{1}{r^2}\frac{\partial}{\partial r}(r^2 U_r) = 0. \tag{A.9}$$

From Eq.(A.9), $\left(\frac{\partial U_r}{\partial r} + 2\frac{U_r}{r}\right) = C_1$, and therefore the general solution of the equation (A.9) in the particle core ("c") and shell ("s") is:

$$U_r^c = C_1 r, \quad U_r^s = C_2 r + \frac{C_3}{r^2}. \tag{A.10a}$$

The strain tensor components are $u_{rr}^{c,s} = \frac{\partial U_r^{c,s}}{\partial r}$ and $u_{\theta\theta}^{c,s} = u_{\phi\phi}^{c,s} = \frac{U_r^{c,s}}{r}$, and their explicit form is

$$u_{rr}^c = u_{\theta\theta}^c = u_{\phi\phi}^c = C_1, \quad u_{rr}^s = C_2 - 2\frac{C_3}{r^3}, \quad u_{\theta\theta}^s = u_{\phi\phi}^s = C_2 + \frac{C_3}{r^3}. \tag{A.10b}$$

Substituting the solution (A.10) into Hooke's law relating the elastic stress and strain tensors, $\sigma_{ij}^{c,s}$ and $u_{ij}^{c,s}$, we obtain the following expressions for radial stresses:

$$\sigma_{rr}^c = \frac{C_1 - Q_c P_3^2 - Z_c P_3^4}{s_{11}^c + 2s_{12}^c + W_c P_3^2} \approx \frac{C_1}{s_{11}^c + 2s_{12}^c} - \frac{Q_c P_3^2 + Z_c P_3^4}{s_{11}^c + 2s_{12}^c}, \tag{A.11a}$$

$$\sigma_{rr}^s = \frac{C_2 - w_s - Q_s P^2}{s_{11}^s + 2s_{12}^s} - 2\frac{C_3}{r^3}\frac{1}{s_{11}^s - s_{12}^s} \approx \frac{C_2 - w_s}{s_{11}^s + 2s_{12}^s} - 2\frac{C_3}{r^3}\frac{1}{s_{11}^s - s_{12}^s}. \tag{A.11b}$$

Here $s_{ij}^c$ are elastic compliances, $Q_c = (Q_{11}^c + 2Q_{12}^c)/3$, $Z_c = (Z_{111}^c + 2Z_{122}^c)/3$, and $W_c = (W_{111}^c + 2W_{112}^c + 2W_{123}^c + 4W_{122}^c)/3$ are isotropic parts of the linear and nonlinear electrostriction tensors of the core; $s_{ij}^s$ are elastic compliances, $Q_s = (Q_{11}^s + 2Q_{12}^s)/3$ is an isotropic part of the linear electrostriction tensor, and $w_s$ are the Vegard strains of the shell. In Eq.(A.11) we assume that the electric field and polarization in the core are homogeneous and directed along the polar axis $X_3$. However, an inhomogeneous stray electric field can exist in the shell, and therefore we consider the total polarization, $P^2 = P_1^2 + P_2^2 + P_3^2$, of the shell. Since the screening length $\lambda$ of the shell is small (less than 1 nm) we can neglect the stray field, and thus omit the electrostriction term, $Q_s P^2$, in the approximate equality in Eq.(11b).

The boundary conditions to Eq.(A.9) are the continuity of radial elastic displacement and normal stress at core-shell interface, $r = R_c$,

$$u_r^c(R_c) = u_r^s(R_c), \quad \sigma_{rr}^c(R_c) = \sigma_{rr}^s(R_c), \tag{A.12a}$$

and the condition of a fixed pressure/tension at the shell surface, $r = R_s$,

$$\sigma_{rr}^s(R_s) = -p. \tag{A.12b}$$

where $p$ is an external pressure or tension. The application of the boundary conditions (A.12) to the solution (A.10)-(A.11) yields the system of equations for the constants $C_i$:

$$C_1 R_c = C_2 R_c + \frac{C_3}{R_c^2}, \qquad (A.13a)$$

$$\frac{C_2 - w_s}{s_{11}^S + 2s_{12}^S} - 2\frac{C_3}{R_c^3}\frac{1}{s_{11}^S - s_{12}^S} = \frac{C_1 - Q_c P_3^2 - Z_c P_3^4}{s_{11}^C + 2s_{12}^C + W_c P_3^2}, \qquad (A.13b)$$

$$\frac{C_2 - w_s}{s_{11}^S + 2s_{12}^S} - 2\frac{C_3}{R_S^3}\frac{1}{s_{11}^S - s_{12}^S} = -p. \qquad (A.13c)$$

For the sake of simplicity below we assume that the elastic compliances of the core and the shell are the same: $s_{11}^S = s_{11}^C = s_{11}$ and $s_{12}^S = s_{12}^C = s_{12}$. In this case, the solution of Eqs.(A.13) is:

$$C_1 = Q_c P_3^2 + Z_c P_3^4 - \frac{s_{11}+2s_{12}}{s_{11}+s_{12}}\frac{2(R_s^3-R_c^3)}{3R_s^3}(Q_c P_3^2 + Z_c P_3^4 - w_s) - (s_{11}+2s_{12})p, \quad (A.14a)$$

$$C_2 = w_s + \frac{s_{11}+2s_{12}}{s_{11}+s_{12}}\frac{2R_c^3}{3R_s^3}(Q_c P_3^2 + Z_c P_3^4 - w_s) - (s_{11}+2s_{12})p, \qquad (A.14b)$$

$$C_3 = \frac{s_{11}-s_{12}}{s_{11}+s_{12}}\frac{R_c^3}{3}(Q_c P_3^2 + Z_c P_3^4 - w_s). \qquad (A.14c)$$

The strain tensor components are:

$$u_{rr}^c = u_{\theta\theta}^c = u_{\phi\phi}^c = Q_c P_3^2 + Z_c P_3^4 - \frac{s_{11}+2s_{12}}{s_{11}+s_{12}}\frac{2(R_s^3-R_c^3)}{3R_s^3}(Q_c P_3^2 + Z_c P_3^4 - w_s) - (s_{11}+2s_{12})p, \qquad (A.15a)$$

$$u_{rr}^s = w_s + \left(\frac{s_{11}+2s_{12}}{s_{11}+s_{12}}\frac{2R_c^3}{3R_s^3} - \frac{s_{11}-s_{12}}{s_{11}+s_{12}}\frac{2R_c^3}{3r^3}\right)(Q_c P_3^2 + Z_c P_3^4 - w_s) - (s_{11}+2s_{12})p, \quad (A.15b)$$

$$u_{\theta\theta}^s = u_{\phi\phi}^s = w_s + \left(\frac{s_{11}+2s_{12}}{s_{11}+s_{12}}\frac{2R_c^3}{3R_s^3} + \frac{s_{11}-s_{12}}{s_{11}+s_{12}}\frac{R_c^3}{3r^3}\right)(Q_c P_3^2 + Z_c P_3^4 - w_s) - (s_{11}+2s_{12})p. \quad (A.15c)$$

Non-trivial stress components in the core are

$$\sigma_{rr}^c = \sigma_{\theta\theta}^c = \sigma_{\phi\phi}^c = -p - \frac{2(R_s^3-R_c^3)}{3R_s^3(s_{11}+s_{12})}(Q_c P_3^2 + Z_c P_3^4 - w_s). \qquad (A.16)$$

For the case $p = 0$, Eq.(A.16) is simplified as:

$$\sigma_{rr}^c = \sigma_{\theta\theta}^c = \sigma_{\phi\phi}^c = -\frac{2(R_s^3-R_c^3)}{3R_s^3}\frac{Q_c P_3^2 + Z_c P_3^4 - w_s}{s_{11}+s_{12}} \approx -2\frac{\Delta R}{R_s}\frac{Q_c P_3^2 + Z_c P_3^4 - w_s}{s_{11}+s_{12}}, \qquad (A.17)$$

where $\Delta R$ is the shell thickness. The approximate equality is valid for thin shells, $(R_s - R_c) \ll R_c$. The nondiagonal stresses are absent, $\sigma_{r\theta}^c = \sigma_{r\phi}^c = \sigma_{\phi\theta}^c = 0$.

Note, that in the important case of different elastic compliances of the core and the shell, the rather cumbersome solution of Eqs.(A.13) can be found. Of particular interest is the stress in the core, which is given by the following expression:

$$\sigma_{rr}^c = \sigma_{\theta\theta}^c = \sigma_{\phi\phi}^c = \frac{-2(R_s^3-R_c^3)(Q_c P_3^2 + Z_c P_3^4 - w_s) - 3R_s^3(s_{11}^S+s_{12}^S)p}{R_s^3(2s_{11}^C + 4s_{12}^C + 2W_c P_3^2 + s_{11}^S - s_{12}^S) - 2R_c^3(s_{11}^C + 2s_{12}^C + W_c P_3^2 - s_{11}^S - 2s_{12}^S)}. \qquad (A.18a)$$

The corresponding free energy of the core-shell nanoparticle is

$$G = \left[\alpha + \frac{1}{\varepsilon_0(\varepsilon_b + 2\varepsilon_s + R_c/\lambda)}\right]\frac{P_3^2}{2} + \beta\frac{P_3^4}{4} + \gamma\frac{P_3^6}{6} + \delta\frac{P_3^8}{8} +$$

$$\frac{3(R_s^3-R_c^3)(Q_c P_3^2 + Z_c P_3^4 - w_s)^2}{R_s^3(2s_{11}^C + 4s_{12}^C + 2W_c P_3^2 + s_{11}^S - s_{12}^S) - 2R_c^3(s_{11}^C + 2s_{12}^C + W_c P_3^2 - s_{11}^S - 2s_{12}^S)}. \qquad (A.18b)$$

Since the condition $s_{11}^S - 2s_{12}^S > 0$ follows from stability of any elastic media, the denominators in Eq.(A.18) are always positive.

The expressions (A.18) are accurate enough to provide a first approximation for the description of the ferroelectric phase of the elastically anisotropic core, when the absolute value of the anisotropic polarization-dependent contribution to the total strain is much smaller than other contributions in Eq.(A.7). The approximation imposes definite conditions on the poorly known (or unknown) anisotropic parts of the nonlinear electrostriction tensors. In order to avoid the complications, the tensors $W_{ijk}$ and $Z_{ijk}$ are regarded elastically isotropic.

### A3. Equation for polarization with renormalized coefficients

After substitution of the solution (A.17) in the LGD equation Eq.(1), the equation of state for the spontaneous polarization yields

$$\left[\alpha + \frac{1}{\varepsilon_0(\varepsilon_b + 2\varepsilon_s + R_c/\lambda)} - (6Q_c\sigma_c + 3W_c\sigma_c^2)\right]P_3 + (\beta - 12Z_c\sigma_c)P_3^3 + \gamma P_3^5 + \delta P_3^7 = 0. \quad (A.19)$$

After substitution of the stress (A.16) into Eq.(A.19) and elementary transformations, the following equation for the spontaneous polarization with renormalized coefficients is obtained:

$$\alpha_R P_3 + \beta_R P_3^3 + \gamma_R P_3^5 + \delta_R P_3^7 + \epsilon_R P_3^9 = 0. \quad (A.20)$$

The renormalized coefficients $\alpha_R, \beta_R, \gamma_R, \delta_R$, and $\epsilon_R$ are given by expressions:

$$\alpha_R = \alpha + \frac{1}{\varepsilon_0(\varepsilon_b + 2\varepsilon_s + R_c/\lambda)} + 6Q_c\left[p - \frac{2(R_s^3 - R_c^3)w_s}{3R_s^3(s_{11}+s_{12})}\right] + 3W_c\left[p - \frac{2(R_s^3 - R_c^3)w_s}{3R_s^3(s_{11}+s_{12})}\right]^2, \quad (A.21a)$$

$$\beta_R = \beta + 6Q_c^2 \frac{2(R_s^3 - R_c^3)}{3R_s^3(s_{11}+s_{12})} + 12Z_c\left[p - \frac{2(R_s^3 - R_c^3)w_s}{3R_s^3(s_{11}+s_{12})}\right] + 6W_cQ_c \frac{2(R_s^3 - R_c^3)}{3R_s^3(s_{11}+s_{12})}\left[p - \frac{2(R_s^3 - R_c^3)w_s}{3R_s^3(s_{11}+s_{12})}\right],$$

(A.21b)

$$\gamma_R = \gamma + 18Q_cZ_c \frac{2(R_s^3 - R_c^3)}{3R_s^3(s_{11}+s_{12})} + 6W_cZ_c \frac{2(R_s^3 - R_c^3)}{3R_s^3(s_{11}+s_{12})}\left[p - \frac{2(R_s^3 - R_c^3)w_s}{3R_s^3(s_{11}+s_{12})}\right] + 3W_cQ_c^2\left(\frac{2(R_s^3 - R_c^3)}{3R_s^3(s_{11}+s_{12})}\right)^2,$$

(A.21c)

$$\delta_R = \delta + 12Z_c^2 \frac{2(R_s^3 - R_c^3)}{3R_s^3(s_{11}+s_{12})} + 6W_cQ_cZ_c\left(\frac{2(R_s^3 - R_c^3)}{3R_s^3(s_{11}+s_{12})}\right)^2, \quad (A.21d)$$

$$\epsilon_R = 3W_cZ_c^2\left(\frac{2(R_s^3 - R_c^3)}{3R_s^3(s_{11}+s_{12})}\right)^2. \quad (A.21e)$$

The shell influence on the coefficients (A.21) yields the ratio $\frac{2(R_s^3 - R_c^3)}{3R_s^3}$. Also, the condition $W_c > 0$ should be valid for the stability of the solutions of Eq.(A.20).